\documentclass[online]{aa}  

\DeclareRobustCommand{\VAN}[3]{#2}
\let\VANthebibliography\thebibliography
\def\thebibliography{\DeclareRobustCommand{\VAN}[3]{##3}\VANthebibliography}

\usepackage{caption}
\usepackage{subcaption}
\usepackage{mhchem}
\usepackage{graphicx}	

\usepackage{amsmath}	
\usepackage{amssymb}	

\begin{document}
\title{The impact of cosmic-ray attenuation on the carbon cycle emission in molecular clouds}
\titlerunning{Cosmic-ray attenuation \& the carbon cycle}
\authorrunning{Gaches et al.}

\author{Brandt A. L. Gaches
        \inst{1}\fnmsep\inst{2}\thanks{E-mail: gaches@ph1.uni-koeln.de}
        \and
        Thomas G. Bisbas
        \inst{1}
        \and 
        Shmuel Bialy
        \inst{3}}
\institute{
I. Physikalisches Institut, Universit\"{a}t zu K\"{o}ln, Z\"{u}lpicher Stra{\ss}e 77, 50937, K\"{o}ln, Germany
\and
Center of Planetary Systems Habitability, The University of Texas at Austin,  USA
\and
Department of Astronomy, University of Maryland, College Park, MD 20742, USA
}

\date{Accepted XXX. Received YYY; in original form ZZZ}

\abstract{Observations of the emission of the carbon cycle species (C, C$^+$, CO) are commonly used to diagnose gas properties in the interstellar medium but are significantly sensitive to the cosmic-ray ionization rate. The carbon-cycle chemistry is known to be quite sensitive to the cosmic-ray ionization rate, $\zeta$, controlled by the flux of low-energy cosmic rays which get attenuated through molecular clouds. However, astrochemical models commonly assume a constant cosmic-ray ionization rate in the clouds.}
{We investigate the effect of cosmic-ray attenuation on the emission of carbon cycle species from molecular clouds, in particular the [CII] 158 $\mu$m, [CI] 609$\mu$m and CO (J = 1 - 0) 115.27~GHz lines.}
{We use a post-processed chemical model of diffuse and dense simulated molecular clouds and quantify the variation in both column densities and velocity integrated line emission of the carbon cycle with different cosmic-ray ionization rate models.}
{We find that the abundances and column densities of carbon cycle species is significantly impacted by the chosen cosmic-ray ionization rate model: no single constant ionization rate can reproduce the abundances modelled with an attenuated cosmic-ray model. Further, we show that constant ionization rate models fail to simultaneously reproduce the integrated emission of the lines we consider, and their deviations from a physically derived cosmic-ray attenuation model is too complex to be simply corrected. We demonstrate that the two clouds we model exhibit a similar average $A_{\rm V, eff}$ -- $n_{\rm H}$ relationship, resulting in an average relation between the cosmic-ray ionization rate and density $\zeta(n_{\rm H})$.}
{We conclude by providing a number of implementation recommendations for CRs in astrochemical models, but emphasize the necessity for column-dependent cosmic-ray ionization rate prescriptions. 
}


\maketitle

\section{Introduction}
Molecular clouds are flooded by energetic, charged particles, called cosmic rays (CRs), accelerated in shocked environments throughout the galaxy \citep{grenier2015, padovani2020}. In dense molecular gas, well-shielded from ultraviolet radiation, CRs are the dominant source of ionization. In particular, the ionization is dominated by protons with energies between 1 MeV -- 1 GeV and secondary electrons produced by primary protons \citep{dalgarno2006}. 

Through the CR ionization of \ce{H2} and formation of \ce{H3+}, subsequent proton-transfer reactions lead to a diverse zoo of molecular chemistry e.g., \ce{HCO+}, CO, \ce{H2O} and \ce{NH3}, as well as initiating the deuteration process \citep[see e.g.][]{Tielens2005, dalgarno2006, bayet2011, caselli2012, indriolo2013, bialy2015, bisbas2015, gaches2019a}. CR ionization also acts as a source of heating in dense gas. Thus, the CR ionization rate (CRIR), denoted $\zeta$ (units H$_2$ ionizations per sec), is one of the most fundamental parameters in astrochemical modeling.

Despite the importance of the CRIR, it is often treated simply with the assumption of a single constant rate. However, low-energy CRs rapidly lose energy as they propagate through molecular gas through Coulomb interactions, ionizations, and pion production \citep{schlickeiser2002, padovani2009}.

There have been several parameterizations of the attenuation of the CRIR as a function of the hydrogen-nuclei column density, hereafter denoted $\zeta(N)$ \citep[e.g.][]{padovani2009, morlino2015, schlickeiser2016, padovani2018, phan2018, ivlev2018, silsbee2019}. Several astrochemical models have included such prescriptions in one-dimensional astrochemical models \citep[e.g.][]{rimmer2012, grassi2019, owen2021, redaelli2021}. While including CR transport (at least high-energy CR transport) has started to become more standard in galactic-scale simulations, to the authors knowledge the inclusion of CR energy losses (and thus a CRIR gradient) has yet to be included in three-dimensional astrochemical models.

Carbon can be found in three major phases in the ISM; the ionized (C$^+$), the atomic (C) and the molecular in the form of CO. Their spectral line emission is of significant importance in controlling the gas temperature and the overall thermal balance in the ISM \citep{hollenbach1999}. They can be also used to determine the ISM environmental parameters and identify the conditions that may lead to star-formation. For example, the emission of $^{12}$CO lines is connected with the presence of molecular gas. The particular $J=1-0$ transition at $115.27\,{\rm GHz}$ and with an energy separation between the two levels of $h\nu/k_{\rm B}\sim5.5\,{\rm K}$, is commonly used to infer to the column density of H$_2$ via the so-called `$X_{\rm CO}-$factor' \citep[see][for a review]{bolatto2013}. Combining all the different CO transitions together into a Spectral Line Energy Distribution (SLED) diagram can reveal much of the ISM conditions that exist in the observed objects. The two emission lines of atomic carbon $^{3}P_1\rightarrow {^3}P_0$ at $609\,\mu$m (hereafter referred to simply as [CI]) and $^{3}P_2\rightarrow {^3P}_1$ at $370\,\mu$m have been also proposed to probe the H$_2$-rich gas \citep{Papadopoulos2004,offner2014,gaches2019b,bisbas2021} in both local \citep[e.g.][]{Lo2014} and extragalactic systems \citep[e.g.][]{Zhang2014,Bothwell2017}. Finally, the particular line of C$^+$ at $158\,\mu$m (hereafter referred to as [CII]) is frequently connected to the presence of warm and ionized gas due to the proximity of its ionization potential (11.2~eV) to that of hydrogen (13.6~eV), although a large fraction of it may originate from molecular gas \citep{Accurso2017,Franeck2018}. 

Recent studies \citep[][]{Meijerink2011, bialy2015, gaches2019a, bisbas2015, bisbas2017b, bisbas2021} have shown that cosmic rays play an immense role in determining the transitions between each carbon cycle phase. Through reactions of CO with He$^+$, the latter of which is a direct consequence of the presence of elevated cosmic-ray energy densities in UV-shielded regions, the abundances of both C$^+$ and C are increased. On the other hand, the transition between atomic and molecular gas remains little affected by ordinary boosts of CRIR. This effect can create extended molecular regions (with $\langle n_{\rm H}\rangle\lesssim10^3\,{\rm cm}^{-3}$) rich in C and even in C$^+$ if the cosmic-ray ionization rate is $\gtrsim500$ times the average one of Milky Way. Thus modelling as accurately as possible the propagation and the attenuation of cosmic rays as a function of column density, is crucial when studying photodissociation and cosmic-ray dominated regions.

In this study, we include a prescription for $\zeta(N)$ in a post-processed three-dimensional astrochemical model of a molecular cloud. In particular, we investigate the impact of the resulting complex three-dimensional CRIR gradients on the abundances and emission from the carbon cycle species, \ce{C}, \ce{C+}, \ce{CO}. In \S 2, we describe the astrochemical CR dominated region (CRDR) model and our prescription for $\zeta(N)$. In \S 3, we discuss the results of our model analysis. We conclude in \S 4.

\section{Methods}

We use the density and velocity distributions of the `dense' and `diffuse' cloud models presented in \citet{bisbas2021}, which are sub-regions taken from the \citet{Wu17} MHD simulations. Each sub-region is a cube of side length $L = 13.88$~pc, resolved with $112^3$ uniform cells. The `dense' cloud has a total mass, $M_{\rm tot}=5.9\times10^4\,{\rm M}_{\odot}$, a mean H-nucleus number density $\langle n_{\rm H} \rangle \equiv M_{\rm tot}/(L^3 \mu m_H) \sim 640\,{\rm cm}^{-3}$, where $m_H$ is the mass of the hydrogen nucleus, and $\mu=1.4$ is the mean particle mass. The mean (observed) H-nucleus column density is $\langle N_{\rm tot} \rangle = \langle n_{\rm H} \rangle L = 2.7 \times10^{22}\,{\rm cm}^{-2}$.  The `diffuse' cloud has a total mass, $M_{\rm tot}=1.9\times10^4\,{\rm M}_{\odot}$, mean H-nucleus number density $\langle n_{\rm H} \rangle = 210\,{\rm cm}^{-3}$ and a mean column density, $\langle N_{\rm tot} \rangle = 9.0\times10^{21}\,{\rm cm}^{-2}$. We focus our primary results on the `dense' cloud, although we note similarities with the `diffuse' model.

We compute the atomic and molecular abundances, $x(i) = n({\rm i})/n_{\rm H}$\footnote{Hereafter, we refer to species abundances and densities through parentheses, and to elemental abundances through subscripts.}, for species i, level populations and gas temperature using the publicly available astrochemical code\footnote{https://uclchem.github.io/3dpdr/} {\sc 3d-pdr} \citep{bisbas2012}. We adopt a subset of the UMIST 2012 chemical network \citep{McElroy13}, which consists of 33 species and 330 reactions, and standard ISM abundances at solar metallicity (n$_{\rm He}$/n$_{\rm H}=0.1$, n$_{\rm C}$/n$_{\rm H}$=$10^{-4}$, n$_{\rm O}$/n$_{\rm H}$=$3\times10^{-4}$; \citealt{Cardelli96, Cartledge04, Rollig07}). We use an external isotropic FUV intensity of $G_0=10$ (normalized according to \citealt{habing68}), a metallicity of $Z=1\,{\rm Z}_{\odot}$ and a microturbulent dispersion velocity of $v_{\rm turb}=2\,{\rm km}\,{\rm s}^{-1}$ to be consistent with \citet{bisbas2021}. Our use of these clouds enables us to investigate the impact of cosmic-ray attenuation of more realistic self-generated clouds, even though the simulations were evolved with different physical parameters (e.g. variable FUV radiation and no CR heating). However, since our goal is to investigate the impact of CR attenuation, the primary results of this work will not be greatly changed by a different external FUV flux.

Furthermore, to construct the emission maps of cooling lines we solve the equation of radiative transfer using the approach presented in \citet{Bisbas2017}, with updates presented in \citet{bisbas2021} to account for dust emission/absorption.

Cosmic-ray attenuation is included using the prescription for the CRIR versus hydrogen column density, $\zeta(N)$, from \citet{padovani2018}, such that
\begin{equation}\label{eq:poly}
    \log_{\rm 10} \left( \frac{\zeta(N)}{\rm s^{-1}} \right) = \sum_{k =0}^9 c_k \left[ \log_{\rm 10} \left( \frac{N_{\rm H, eff}}{\rm cm^{-2}} \right) \right]^k,
\end{equation}
where the coefficients, $c_k$\footnote{For implementation into {\sc Fortran} codes, it is heavily advised to use 16 digit precision reals, to avoid numerical inaccuracy after the logarithm is exponentiated.}, are taken from Table F.1 of \citet{padovani2018}, shown also in Table \ref{tab:crparams}, and $N$ is the total hydrogen column density `seen' by the cosmic ray. We use the $\mathcal{H}$ model, which reproduces the AMS-02 high-energy CR spectrum \citep{aguilar2014, aguilar2015} and adjusts the low energy spectrum, below $E < 1$ GeV, to reproduce the CRIR inferred in diffuse gas \citep[e.g.][]{indriolo2012, indriolo2015, neufeld2017}. For the hydrogen column density, we utilize the effective (local) hydrogen column density \citep{glover2010} computed as part of the radiation transfer calculations in {\sc 3d-pdr},
\begin{equation}
    N_{\rm H, eff} = -\frac{1}{2.5} \ln \left ( \frac{1}{\mathcal{N}_{\ell}} \sum_{i=1}^{\mathcal{N}_{\ell}} e^{-2.5 N_{\rm H,i}} \right ),
\end{equation}
where we sum over the $\mathcal{N}_{\ell}=12$ HEALPix rays of the $\ell=0$ level onto which the hydrogen number density is projected and summed along (see \citealt{bisbas2012} for details). 

\begin{table}[h!]
    \caption{The $c_k$ coefficients for Equations \ref{eq:poly} and \ref{eq:polyn}. The coefficients, $c_k$, are taken from Table F.1 from \citet{padovani2018}.}
    \label{tab:crparams}
    \centering
    \begin{tabular}{c|c}
        k & $c_k$ \\
        \hline
        0 & $\phantom{-}1.001098610761\times10^7$\\
        1 & $-4.231294690194\times10^6$ \\
        2 & $\phantom{-}7.921914432011\times10^5$ \\
        3 & $-8.623677095423\times10^4$\\
        4 & $\phantom{-}6.015889127529\times10^3$ \\
        5 & $-2.789238383353\times10^2$\\
        6 & $\phantom{-}8.595814402406\times10^0$ \\
        7 & $-1.698029737474\times10^{-1}$\\
        8 & $\phantom{-}1.951179287567\times10^{-3}$ \\
        9 & $-9.937499546711\times10^{-6}$
    \end{tabular}
\end{table}

While the true column density, $N$, should take into account the propagation along potentially twisted magnetic field lines, our assumption allows for a first examination of the impact of CR attenuation in three-dimensions and represents a case where the magnetic field lines are not significantly tangled (valid for $n_{\rm H} \le 10^9$ cm$^{-3}$, see \citealt{padovani2013}). We compare our CR-attenuated CRDR model with four constant-CR models, $\zeta_c = (1, 2,  5, 10) \times 10^{-16}$ s$^{-1}$. The particular $\zeta_c = \zeta_M \equiv 2 \times 10^{-16}$ s$^{-1}$ rate corresponds to the total mass-weighted CRIR in the domain, calculated using the $\zeta(N)$ model:
\begin{equation}
    \zeta_M =  \frac{\int n_{\rm H}(x,y,z) \zeta(x,y,z) dV}{\int n_{\rm H}(x,y,z) dV} \approx 2\times10^{-16}\,{\rm s^{-1}},
\end{equation}
where the integral is performed over the volume, $V$, of the simulation domain.

In this work, we assume that the impinging CR flux is isotropic and free-stream into the cloud. Deviations from these assumptions would incur a locally anisotropic cosmic-ray flux. Cosmic-ray magneto-hydrodynamic simulations of molecular clouds are needed to  elucidate the significance of this anisotropy or post-processing with more sophisticated treatments of cosmic-ray transport (see e.g. \citealt{fitzaxen2021} for this treatment with point sources).

\section{Results and discussion}
\subsection{Gas and cosmic-ray ionization rate distributions}
\begin{figure}
    \centering
    \includegraphics[width=0.5\textwidth]{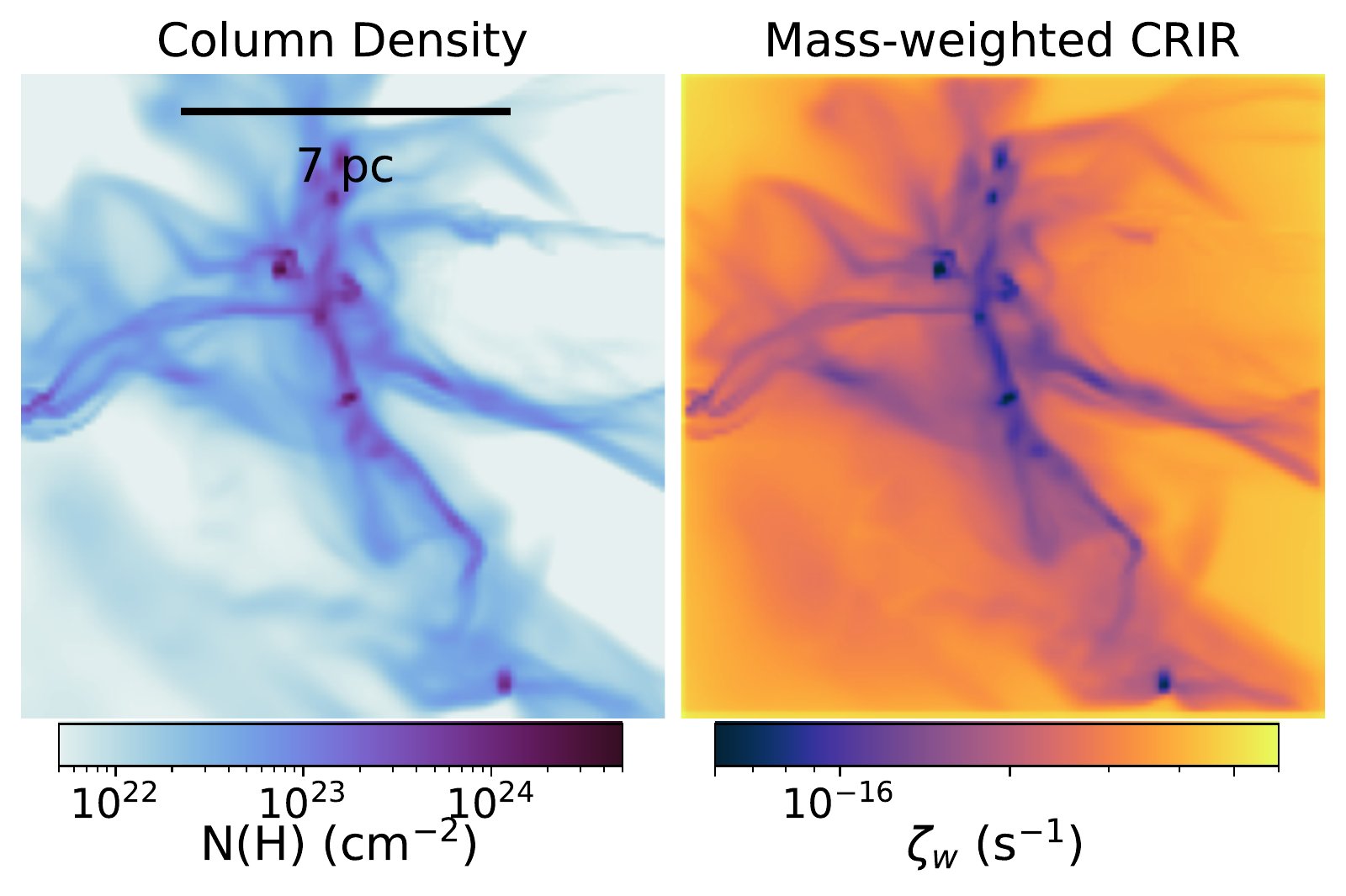}
    \caption{\label{fig:sim_grid} Total gas column density (left) and line-of-sight mass-weighted CRIR (Eq.~\ref{eq:zetam}) (right) for the `dense' (top) and `diffuse' (bottom) cloud models.}
\end{figure}

\begin{figure}
    \centering
    \includegraphics[width=0.5\textwidth]{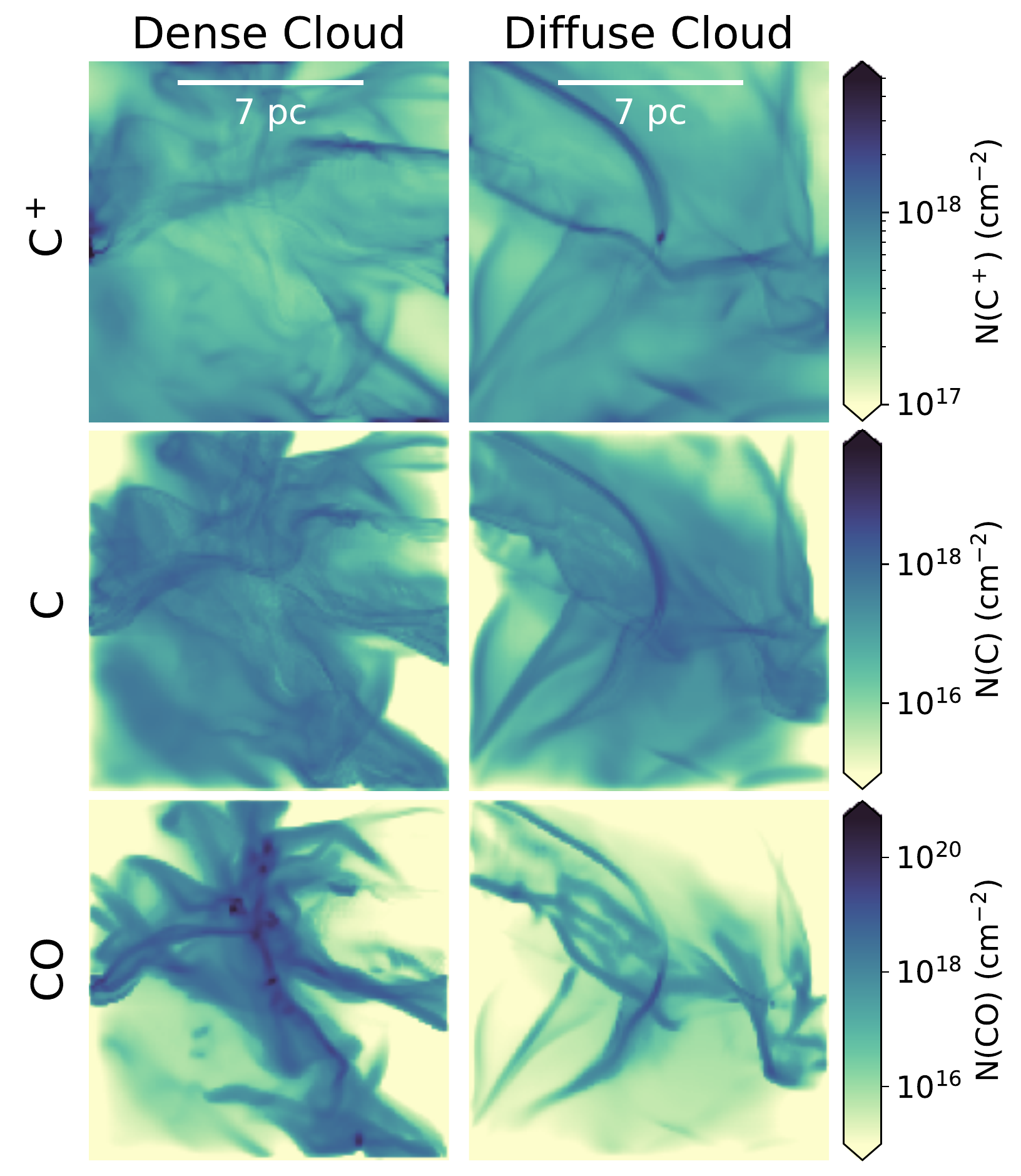}
    \caption{\label{fig:colsum}Summary of the carbon-cycle column densities for the `dense' (top) and `diffuse' (bottom) cloud using the $\zeta(N)$ model.}
\end{figure}

\begin{figure}
    \centering
    \includegraphics[width=0.5\textwidth]{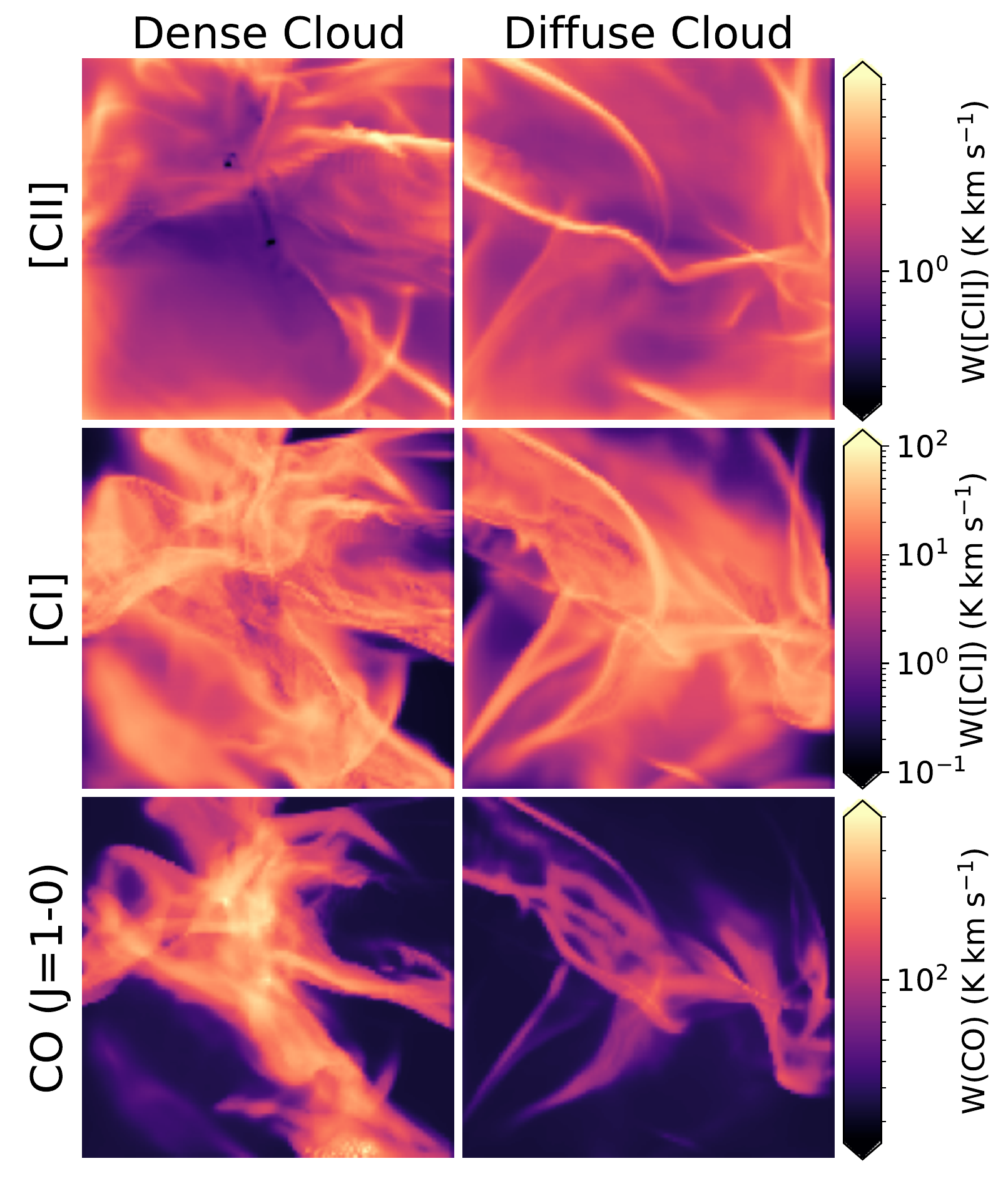}
    \caption{\label{fig:emisum}Summary of the carbon-cycle line-of-sight integrated fluxes for the `dense' (top) and `diffuse' (bottom) cloud using the $\zeta(N)$ model.}
\end{figure}

\begin{figure}
    \centering
    \includegraphics[width=0.5\textwidth]{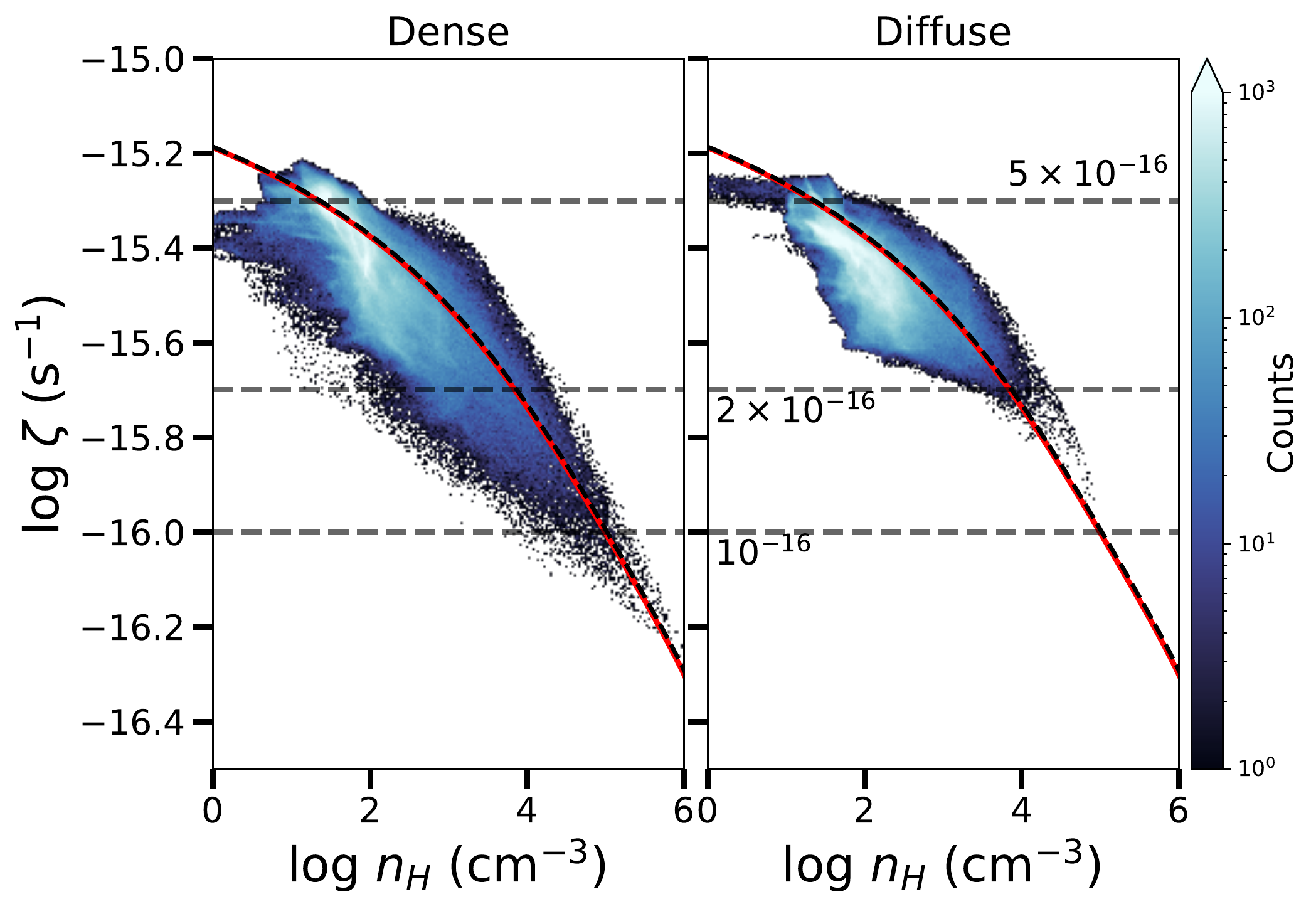}
    \caption{\label{fig:zeta}The local cosmic-ray ionization rate -- density 2D-PDF for the `dense' and `diffuse' clouds. The red line is a parameterization utilizing an $A_{\rm V, eff}$--$n_{\rm H}$ relation (see Eq \ref{eq:avn}) and the black dashed line is the analytic equation by Eq. \ref{eq:polyn}..}
\end{figure}

\begin{figure}
    \centering
    \includegraphics[width=0.5\textwidth]{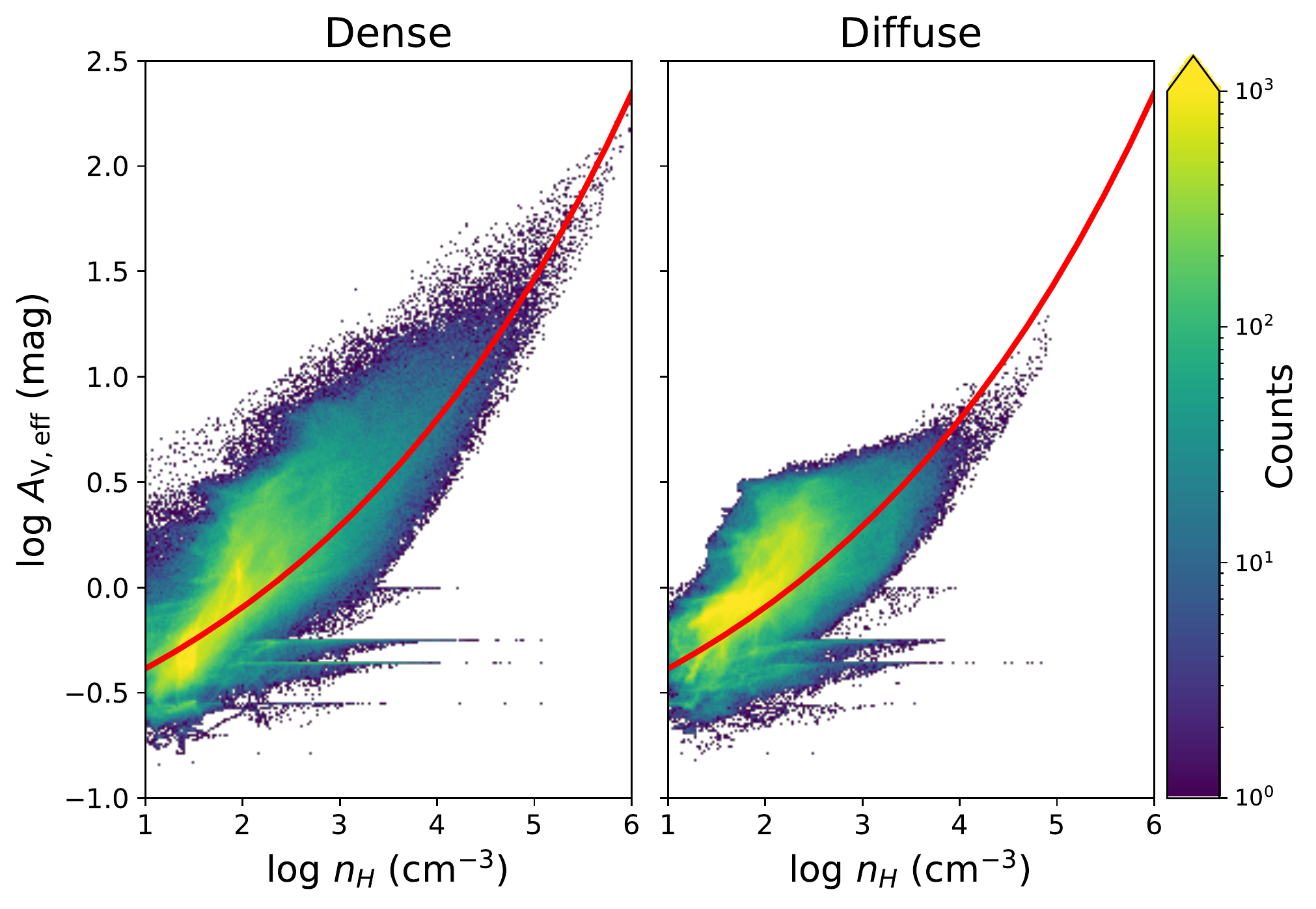}
    \caption{\label{fig:avnef}. The $A_{\rm V, eff}$ -- $n_{\rm H}$ 2D-PDF for the `dense' and `diffuse' clouds. The red-line is the $A_{\rm V, eff}$--$n_{\rm H}$ relation (see Eq \ref{eq:avn}).}
\end{figure}

Figure \ref{fig:sim_grid} shows the gas column densities and the line-of-sight mass-weighted CRIR, 
\begin{equation}\label{eq:zetam}
    \zeta_w(x,y) = \frac{\int n_{\rm H}(x,y,z) \zeta(x,y,z) dz}{\int n_{\rm H}(x,y,z) dz},
\end{equation}
for both `dense' and `diffuse' clouds. We see that there is a strong correlation of the gas column density with the mass-weighted CRIR along the line-of-sight. There is an order of magnitude variation of the CRIR throughout the cloud due to the attenuation. Figure \ref{fig:colsum} shows the column density integrated along the Z axis\footnote{Throughout this paper, projection plots integrate along the Z axis unless otherwise stated.} for the carbon cycle species we consider for the two different clouds. Figure \ref{fig:emisum} summarizes the velocity integrated line emission of [CII]~158$\mu$m, [CI]~609$\mu$m and CO ($J = 1-0$) for both clouds. We will discuss the aforementioned results in more detail below.

Figure \ref{fig:zeta}  shows the 2D-PDF distribution, hereafter referred to as a `phase space', of CRIR and density for the `dense' and `diffuse' clouds. The distribution shows that low-density regions encounter a CRIR an order of magnitude greater than at the highest densities. The overall trends can be explained by a self-gravitating cloud with turbulence induced porosity, such that lower density gas is more likely to be closer to the edge of the domain and dense gas is typically found in more embedded regions in the cloud. The two clouds show remarkable similarities in the distribution functions for the same density range. 

\subsection{The average $\zeta$ -- $n_{\rm H}$ relation}
Here, we discuss the average $\zeta$ -- $n_{\rm H}$ relation in both clouds, and the implications for a more general relation. Numerous previous theoretical investigations have found an average trend between the effective (local) $A_{V,eff}$ and the total H-nucleus number density, $n_{\rm H}$, \citep[see e.g.][]{glover2010, vanloo2013, safranek2017, seifried2017, Bisbas2019, hu2021}. Figure \ref{fig:avnef} shows the $A_{\rm V, eff} - n_{\rm H}$ distribution for the two simulations along with a best fit function from \citet{Bisbas2019}, Bisbas et al. (in prep):
\begin{equation}\label{eq:avn}
    A_{\rm V, eff}(n_{\rm H}) = A_{\rm V,0} \exp \left [a \left ( \frac{n_{\rm H}}{{\rm cm}^{-3}} \right )^{\gamma} \right ]
\end{equation}
where $A_{\rm V,0}=0.05$, $a=1.6$, and $\gamma=0.12$,  and where for typical dust properties
$A_{\rm V, eff}$ and $N_{\rm H}$ are related through $N/A_{\rm V} = 1.6 \times 10^{21}$ cm$^{-2}$ $\equiv f$ \citep[][]{bohlin1978, weingartner2001, Rollig07, rachford2009}.
We find that the best fit function reasonably reproduces the average trends in our distribution functions.
A further  discussion of the $A_{\rm V, eff}$ -- $n_{\rm H}$ relation is deferred to \citet{Bisbas2019} (in particular their Appendix B and Figure B.1) and Bisbas et al. (in prep). 

Combining Eqs.~(\ref{eq:poly}) and (\ref{eq:avn}) we get an analytic equation for the mean relationship of the attenuated CRIR and the density in the gas. We get:
\begin{equation}\label{eq:polyn}
    \log_{\rm 10} \left( \frac{\zeta}{\rm s^{-1}} \right) = \sum_{k =0}^9 c_k \left[ A+B \left(\frac{n_{\rm H}}{\rm cm^{-3}} \right)^{\gamma} \right]^k,
\end{equation}
where $A=\log_{10}(f A_{\rm V,0}) = 19.90$, $B= a \log_{10}\mathrm{e} = 0.69$, $\gamma=0.12$ (Eq.~\ref{eq:avn}), and the $c_k$ coefficients are the same as in Eq.~(\ref{eq:poly}) and are given in Table \ref{tab:crparams}.
This $\zeta-n_{\rm H}$ relation is shown in Figure \ref{fig:zeta} (red-dashed curve).
This simple prescription is able to reproduce the average behavior of the 2D distribution for both clouds.
The universality of the $\zeta-n_{\rm H}$ relation is rooted at the universality of the $A_{\rm V,eff} - n_{\rm H}$ relation, which is found to be remarkably robust across seven orders of magnitude in density, and is seen in various simulation codes (the $A_{\rm V,eff} - n_{\rm H}$ relation does, however, depend on metallicity; see \citealt{hu2021}). 
We note that this trend is likely to hold only in the limit that the CR flux is relatively isotropic, that the transport is dominated by energy-losses and not turbulent diffusion (and thus only dependent on the column density) and that there are no embedded sources of CRs. 

Due to the similarity of the $\zeta$ -- $n_{\rm H}$ phase space between the two clouds and of their $A_{\rm V, eff}$ -- $n_{\rm H}$ phase spaces, we expect the qualitative impact of CR attenuation to be similar. Therefore, in the rest of this paper, we focus only on the `dense' cloud simulation.

\subsection{Impact on the distribution of carbon cycle species}
\begin{figure*}
    \centering
    \includegraphics[trim=75 15 25 15,clip,width=0.95\textwidth]{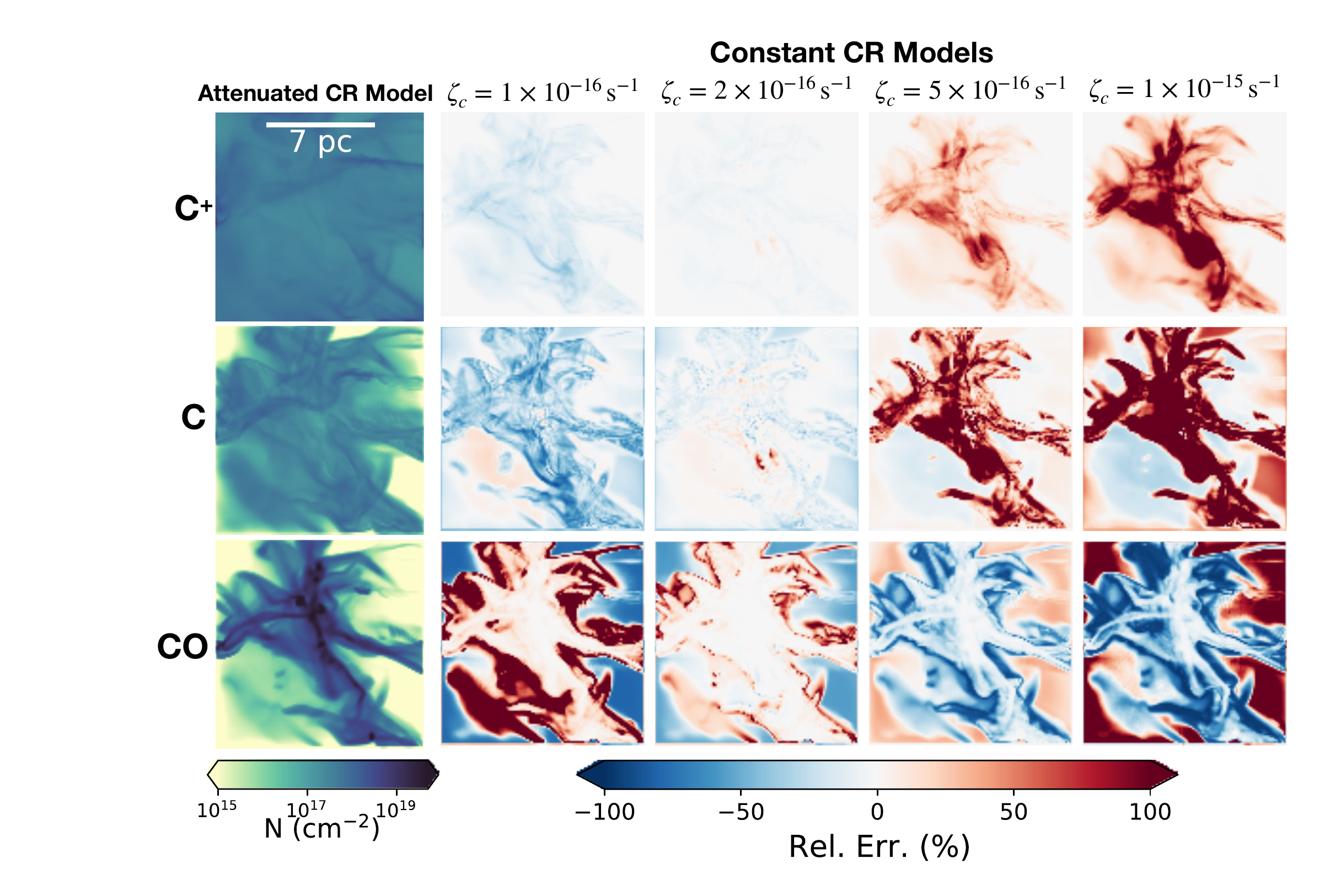}
    \caption{\label{fig:colStamp} First column: Column densities of ionized carbon (top), neutral carbon (middle) and carbon monoxide (bottom) for our non-constant $\zeta(N)$ model. Second to Fourth columns: The relative difference (see Eqn.~\ref{eqn:epsilon}) in column density for a constant CRIR with respect to the $\zeta(N)$ model for each of the aforementioned species. }
\end{figure*}

We will first discuss the impact of CR attenuation on the distribution of \ce{C+}, \ce{C} and \ce{CO}. Figure \ref{fig:colStamp} shows the column densities of \ce{C+}, \ce{C} and \ce{CO} for the $\zeta(N)$ model and the relative differences in the column densities, defined by
\begin{equation}
\label{eqn:epsilon}
    \varepsilon(N_{i}) = \frac{ N_{i, \zeta_c} - N_{i, \zeta(N)} }{N_{i, \zeta(N)}},
\end{equation} 
for a given species, $i$,  with constant CRIR, $\zeta_c$ relative to the $\zeta(N)$ model. We find that the $\zeta_c=5 \times 10^{-16}$ s$^{-1}$ model significantly overproduces \ce{C+} and \ce{C} and under-produces CO, while the $\zeta_c=10^{-16}$ s$^{-1}$ model under-produces \ce{C+} and \ce{C} but overproduces CO. The model with $\zeta_c = 2\times10^{-16}$ s$^{-1}$ has the minimal error. The behavior of CO is interesting; the $\zeta_c = 5\times10^{-16}$ s$^{-1}$ model over produces CO in diffuse regions while under-producing CO in a thin zone surrounding the dense gas structures. This trend is flipped for the $\zeta_c = 2\times10^{-16}$ and $5\times10^{-16}$ s$^{-1}$ models. At high densities, the CO distributions are effectively identical.

\begin{figure*}
    \centering
    \includegraphics[width=0.95\textwidth]{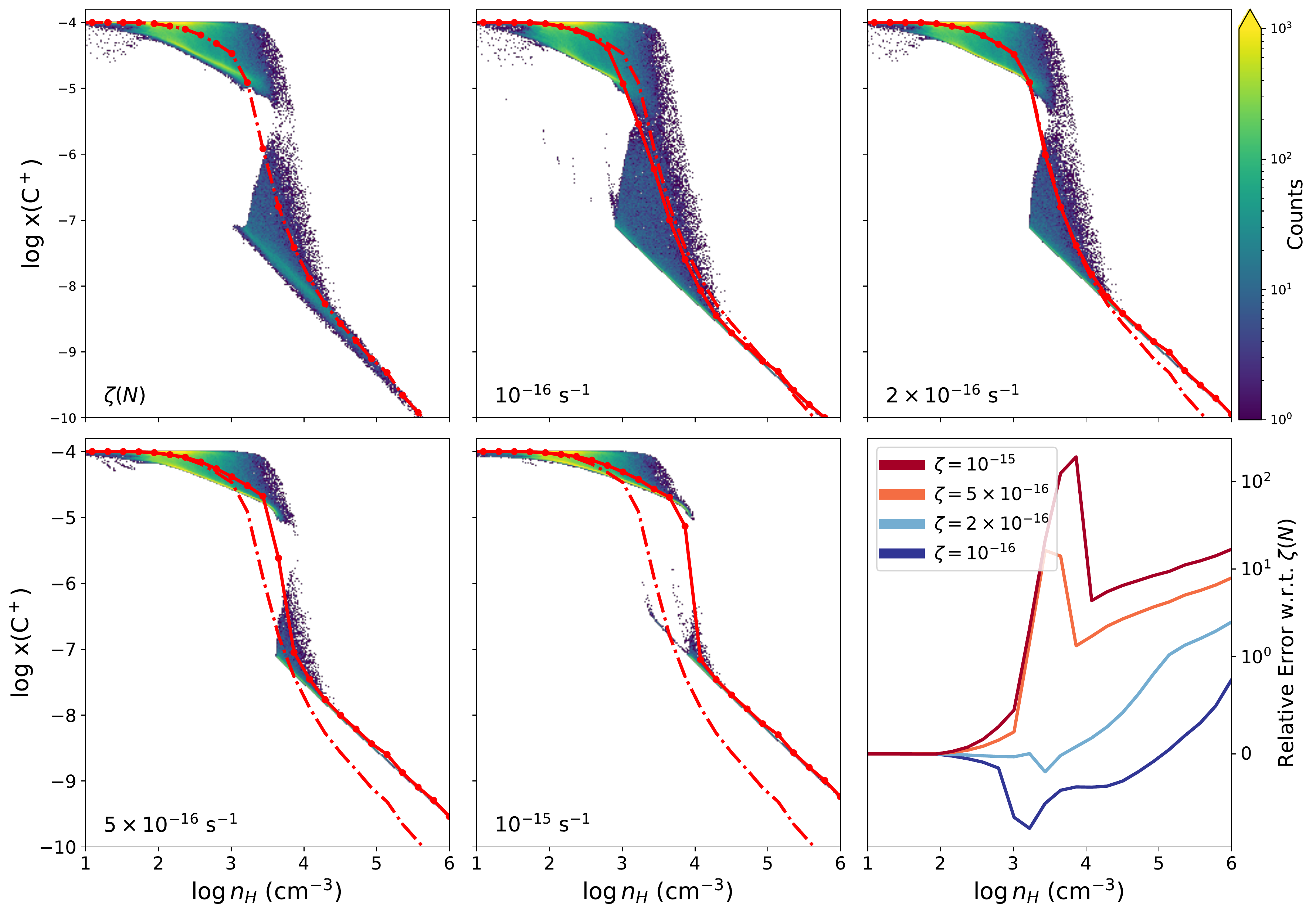}
    \caption{\label{fig:ciiabund} \ce{C+} abundance, x(\ce{C+}) versus hydrogen nuclei density, $n_{\rm H}$. The cosmic-ray model is annotated in each panel. The red lines denote the binned averaged abundance profiles in log-log space. The dashed-dotted lines shows the $\zeta(N)$ abundance profile for comparison. The bottom right panel shows the relative error of the average abundance profiles.}
\end{figure*}

\begin{figure*}
    \centering
    \includegraphics[width=0.95\textwidth]{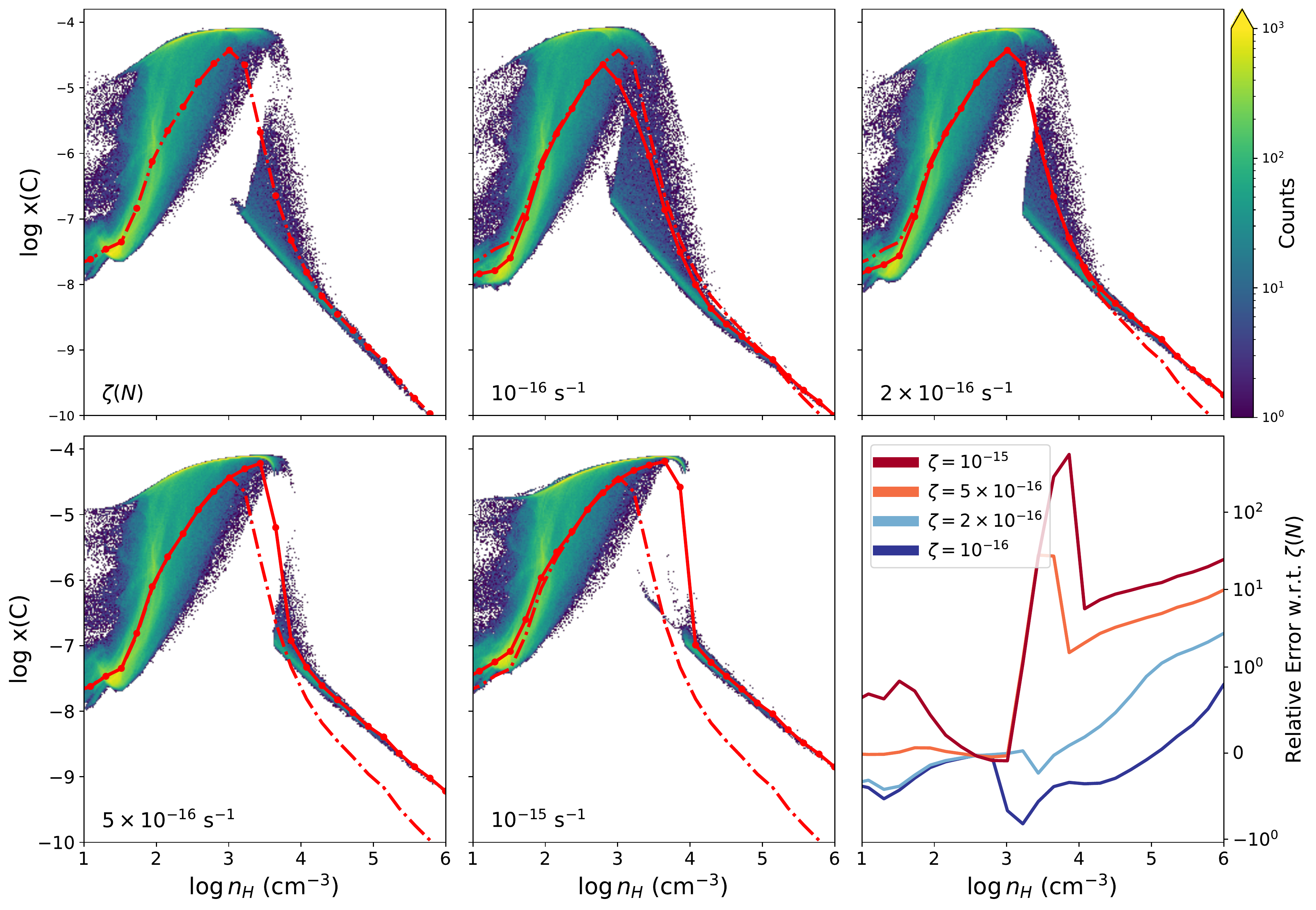}
    \caption{\label{fig:ciabund} As in Figure \ref{fig:ciiabund} for C.}
\end{figure*}

\begin{figure*}
    \centering
    \includegraphics[width=0.95\textwidth]{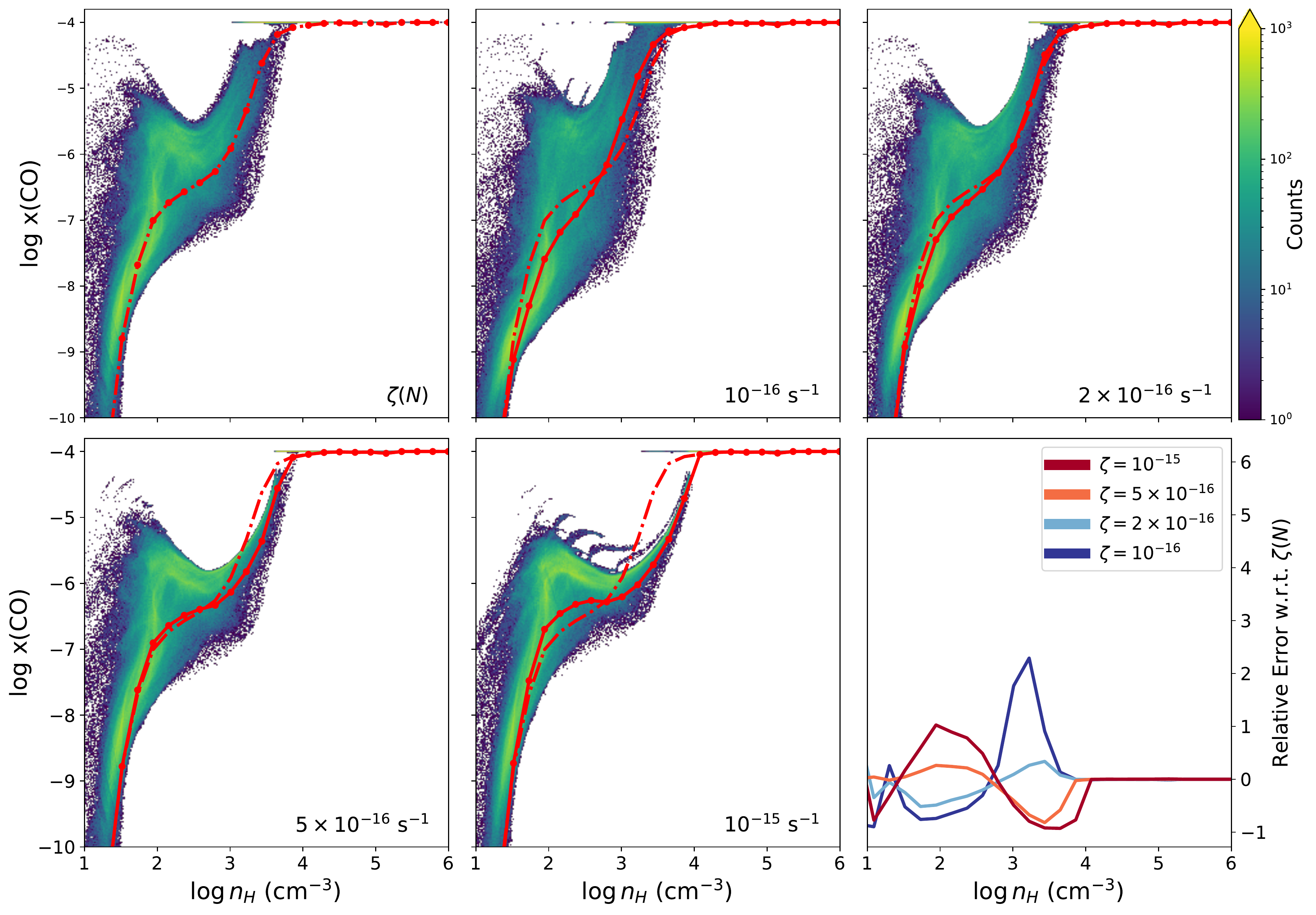}
    \caption{\label{fig:coabund} As in Figure \ref{fig:ciiabund} for CO.}
\end{figure*}

Figures \ref{fig:ciiabund}-\ref{fig:coabund} show the $x({\rm C^+})-n$, $x({\rm C})-n$, $x({\rm CO})-n$ phase spaces, for our various simulations with the bin-averages as red lines. The last panel on the bottom right of each figure shows the linear-relative error in the average abundance profiles for a specific constant CRIR model compared to the $\zeta(N)$ model.

Looking at Figures \ref{fig:ciiabund} -- \ref{fig:coabund}, we can identify several general trends. First, for densities $n_{\rm H} \lesssim 100$ cm$^{-3}$, the abundance of C$^+$ is insensitive to the $\zeta$. This is because gas at these densities is preferentially located near the cloud boundary (following the Eq. \ref{eq:avn}, this gas has $A_{\rm V eff} \leq 0.8$, on average) where carbon photo-ionization is efficient and dominates the production of C$^+$\footnote{The ionization potential of C is $E = 11.3$ eV. Thus, even in regions where the hydrogen is predominantly neutral (i.e., in the bulk of the ISM), carbon may be readily photo-ionized by photons with energies in the range $E=(11.3-13.6)$ eV.}. In this regime, photo-ionization keeps the carbon predominantly in ionized form, with $x({\rm C^+}) \simeq A_C \approx 10^{-4}$, where $A_C$ is the elemental carbon abundance.

Then, as we move to higher densities, we sample regions at higher $A_V$, at which point dust absorption becomes important significantly reducing the efficiency of carbon photo-ionization. At $n_{\rm H} \gg 100$ cm$^{-3}$, CR-driven reactions dominate over photo-ionization and the C$^+$ abundance then depends on $\zeta$. In this regime, as the CRIR increases, the abundances of both C and C$^+$ increase, at the expense of CO. This is because CRs (1) ionize He resulting in the formation of He$^+$, and (2) excite H$_2$ molecules generating an FUV radiation inside the cloud interior \citep{sternberg1987, gredel1989, heays2014, bialy2020}. Both the He$^+$ ions and the FUV cosmic-ray produced FUV radiation result in the destruction of CO molecules and the formation of C and C$^+$. Similarly, for very high CRIRs, the induced FUV radiation can become an important source of C$^+$ through the photo-ionization of C. Thus, the abundances of C and C$^+$ generally increase with $\zeta$. For a thorough discussion of these chemical reactions, see \S 2.3 in \citealt{bialy2015}, see also their \S 5 for analytic scaling relations (in particular, their Eq.~45).

Indeed,  looking at Figs. \ref{fig:ciiabund} -- \ref{fig:ciabund}, we see that within this high-density regime, the C and C$^+$ abundances are always higher for the two models with high  CRIRs ($\zeta_c = (5, 10) \times 10^{-16}$ s$^{-1}$), compared to the $\zeta(N)$ model (at these densities the $\zeta(N)$ model gives a CRIR that is typically below $5 \times 10^{-16}$ s$^{-1}$; see Fig. \ref{fig:zeta}).

Looking at the bottom-right panels of Figs. \ref{fig:ciiabund} -- \ref{fig:ciabund}, we see that as the density increases, the relative abundances of C and C$^+$ (in each of the $\zeta_c$ models, relative to the $\zeta(N)$ model) increases with increasing $\zeta_c$. At sufficiently high densities, even the models with $\zeta_c=2 \times 10^{-16}$ s$^{-1}$ and $\zeta_c=10^{-16}$ s$^{-1}$ exhibit a higher C and C$^+$ abundances, compared to the $\zeta(N)$ model. This is because while the CRIR remains constant in the $\zeta_c$ models, it {\it decreases} with $n_{\rm H}$ in the $\zeta(N)$ model, as seen in Fig. \ref{fig:zeta} (see also Eq. \ref{eq:polyn}). Specifically, at the points, $n_{\rm H} \approx 5 \times 10^3$ and $10^5$ cm$^{-3}$, the CRIR in the $\zeta(N)$ model falls below the values of the two constant CRIR models, $\zeta_c=2 \times 10^{-16}$ s$^{-1}$ and $\zeta_c=10^{-16}$ s$^{-1}$, respectively. These are the points where there is a transition from underproduction to over-production of C and C$^+$ in the $\zeta_c=(1,2) \times 10^{-16}$ s$^{-1}$ models (compared to the $\zeta(N)$ model), i.e., the curves in the bottom-right panels cross zero.

Looking at Fig. \ref{fig:coabund}, we see that while the region where the CO abundance is sensitive to $\zeta$ is limited to low and intermediate densities, $n_{\rm H} \lesssim 10^4$ cm$^{-3}$, whereas for $n_{\rm H} \approx 10^4$ cm$^{-3}$ CO becomes essentially independent of $\zeta$. This is because at even though CRs drive the destruction processes of CO, at these high densities, the CO formation rate strongly exceeds the destruction rate (the CO formation-to-destruction rate ratio increases with $n_{\rm H}/\zeta$). Thus, the carbon becomes predominantly in the form of CO, and the CO abundance saturates at the elemental carbon abundance, $x_{\rm CO} \approx A_{\rm C}$.

\subsection{Variations in the observed emission}
\begin{figure*}
    \centering
    \includegraphics[trim=0 15 15 15, clip,width=0.95\textwidth]{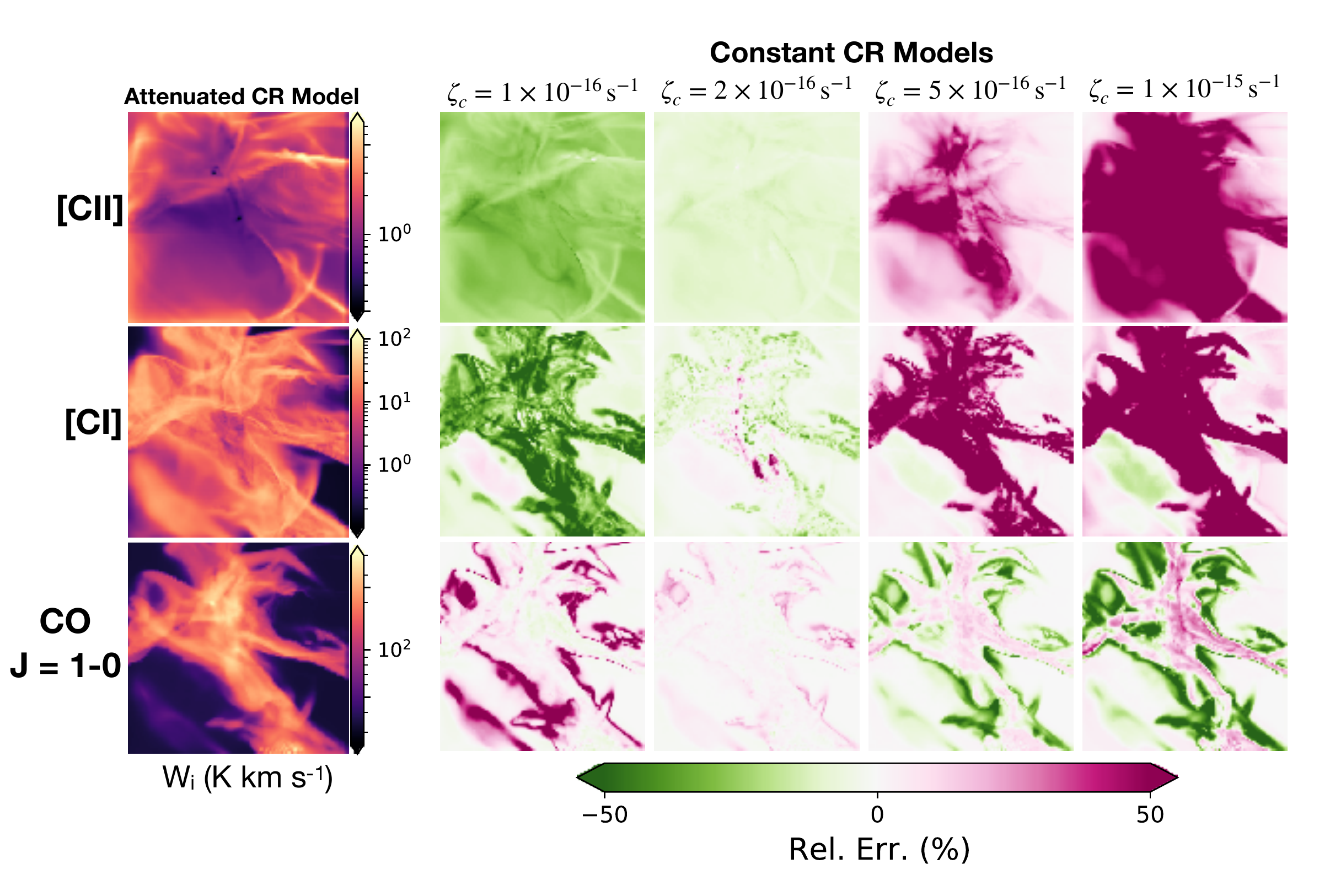}
    \caption{\label{fig:intStamp} First column: Velocity integrated emission of the [C{\sc ii}] 158 $\mu$m (top), [C{\sc i}] 609 $\mu$m (middle) and CO (1-0) (bottom) emission in units of (K ${\rm km}\,{\rm s}^{-1}$). Second to Fourth columns: Relative emission for the different constant cosmic-ray ionization rate models with respect to the $\zeta(N)$ model.}
\end{figure*}

\begin{figure*}
    \centering
    \begin{subfigure}[b]{0.3075\textwidth}
        \centering
        \includegraphics[width=\textwidth]{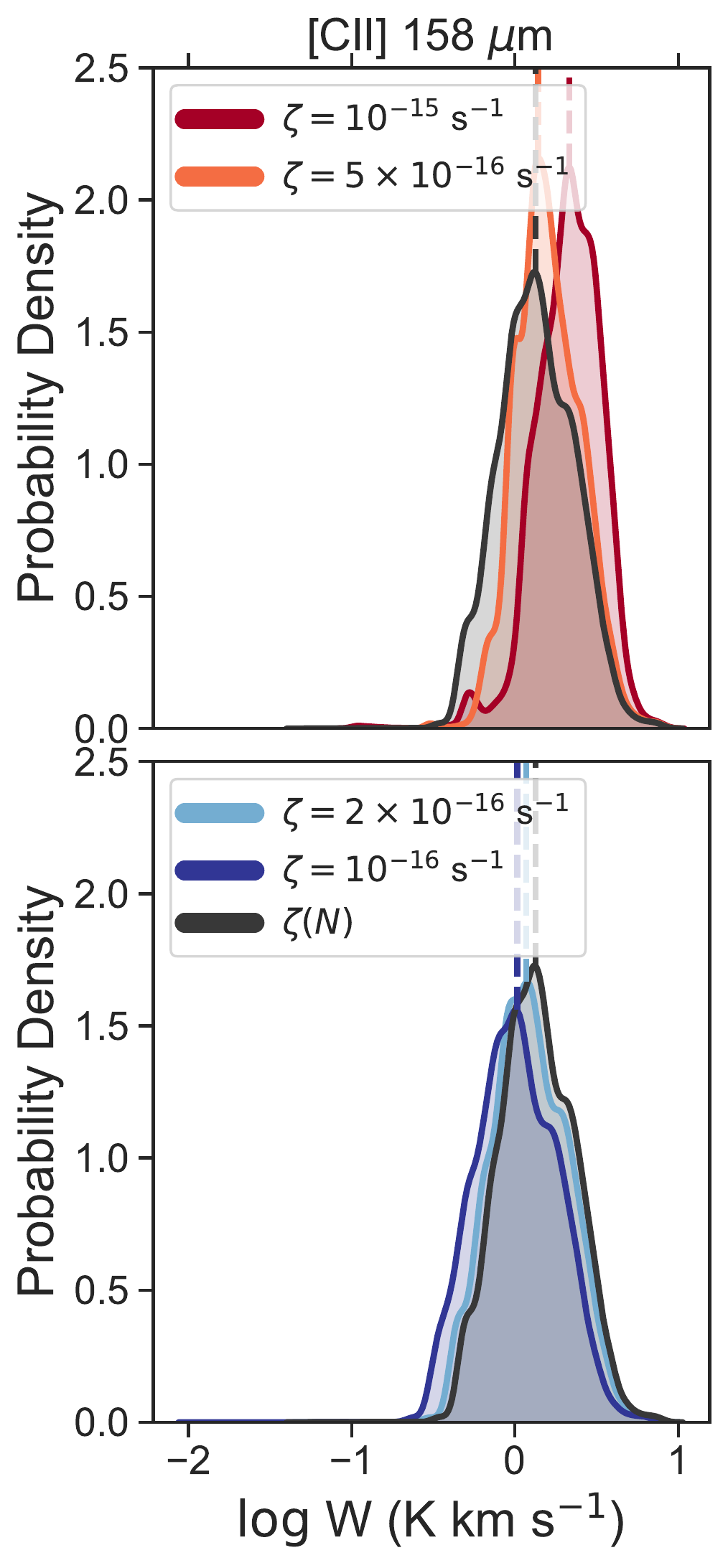}
        \label{fig:5a}
    \end{subfigure}
    \begin{subfigure}[b]{0.283\textwidth}
        \centering
        \includegraphics[width=\textwidth]{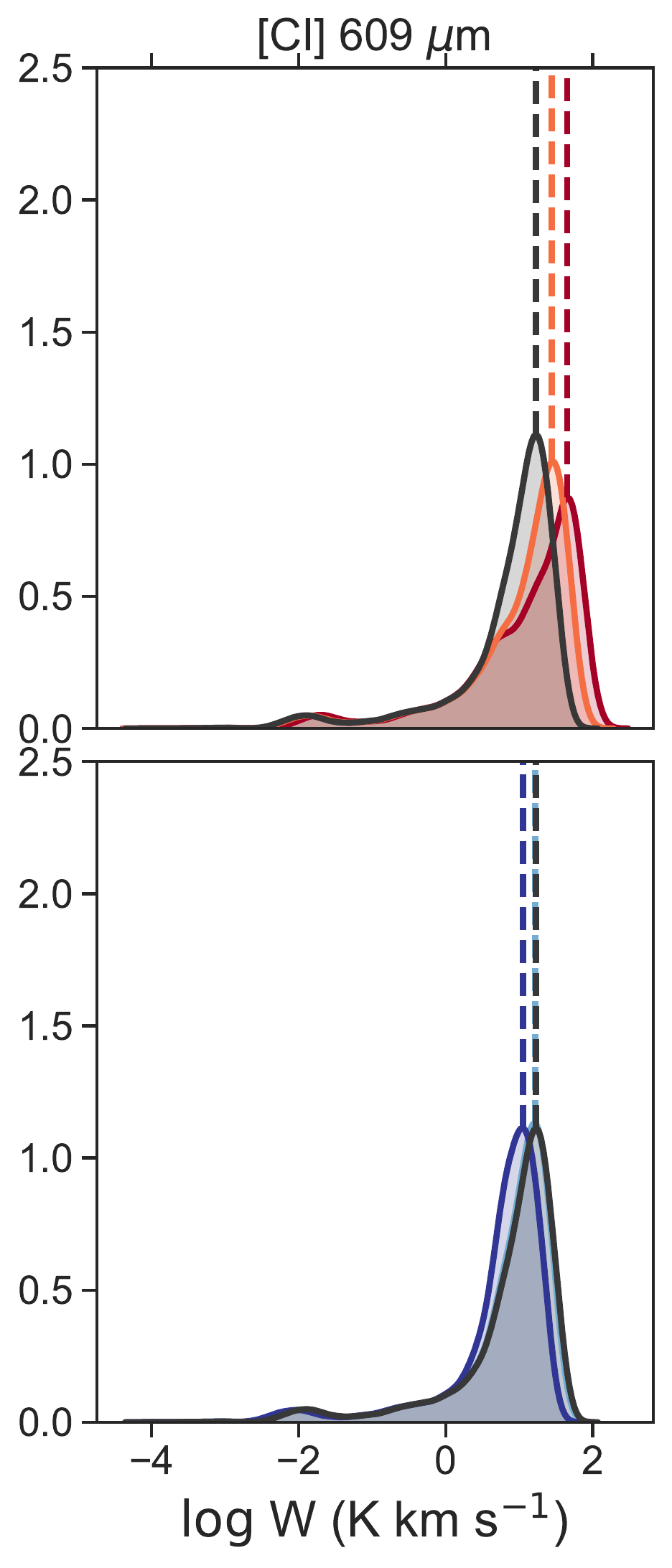}
        \label{fig:5b}
    \end{subfigure}
    \begin{subfigure}[b]{0.283\textwidth}
        \centering
        \includegraphics[width=\textwidth]{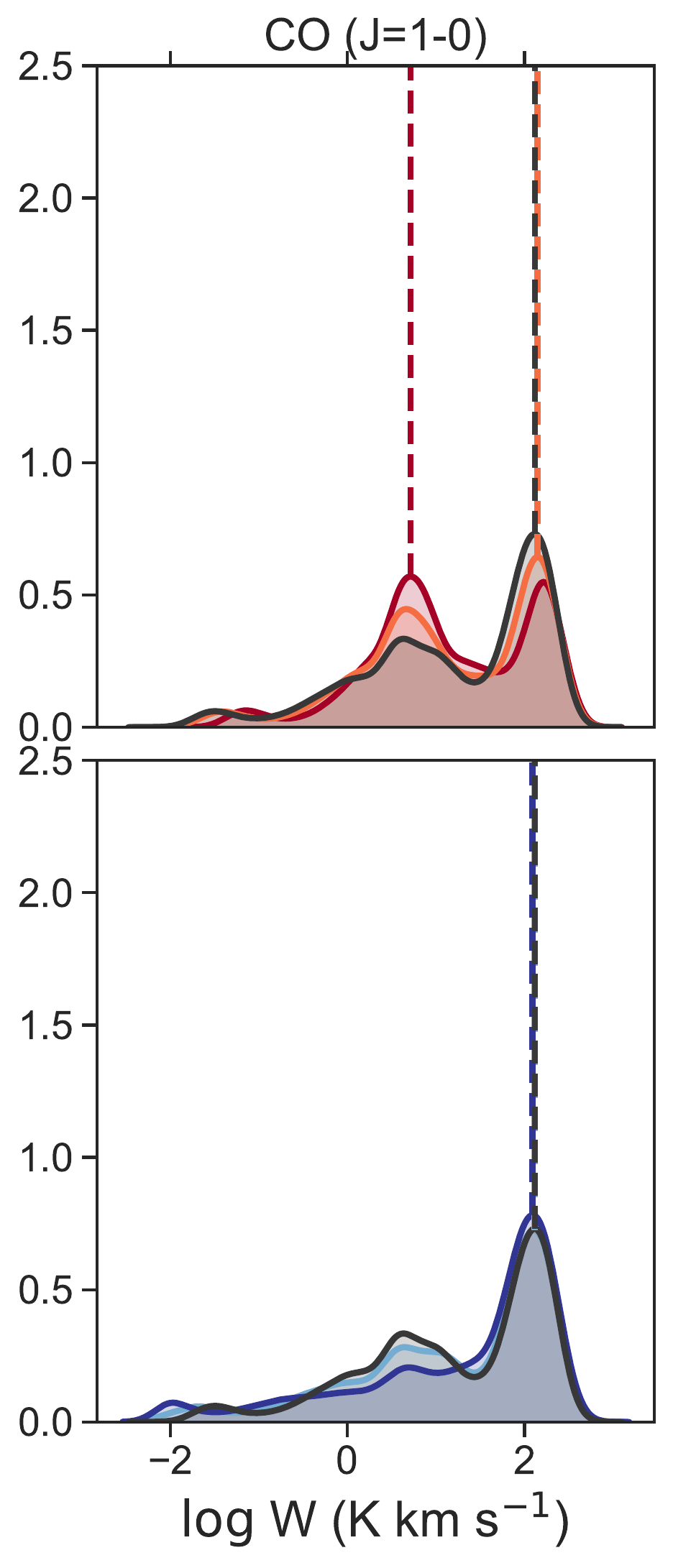}
        \label{fig:5c}
    \end{subfigure}
    \caption{\label{fig:flxDist} PDFs of the velocity integrated emission of [C{\sc ii}] 158 $\mu$m (right), [C{\sc i}] 609$\mu$m (center),  and CO (1-0) (right) for the different ionization rate models.The vertical lines highlight the peak emission.}
\end{figure*}

\begin{figure}
    \centering
    \includegraphics[width=0.45\textwidth]{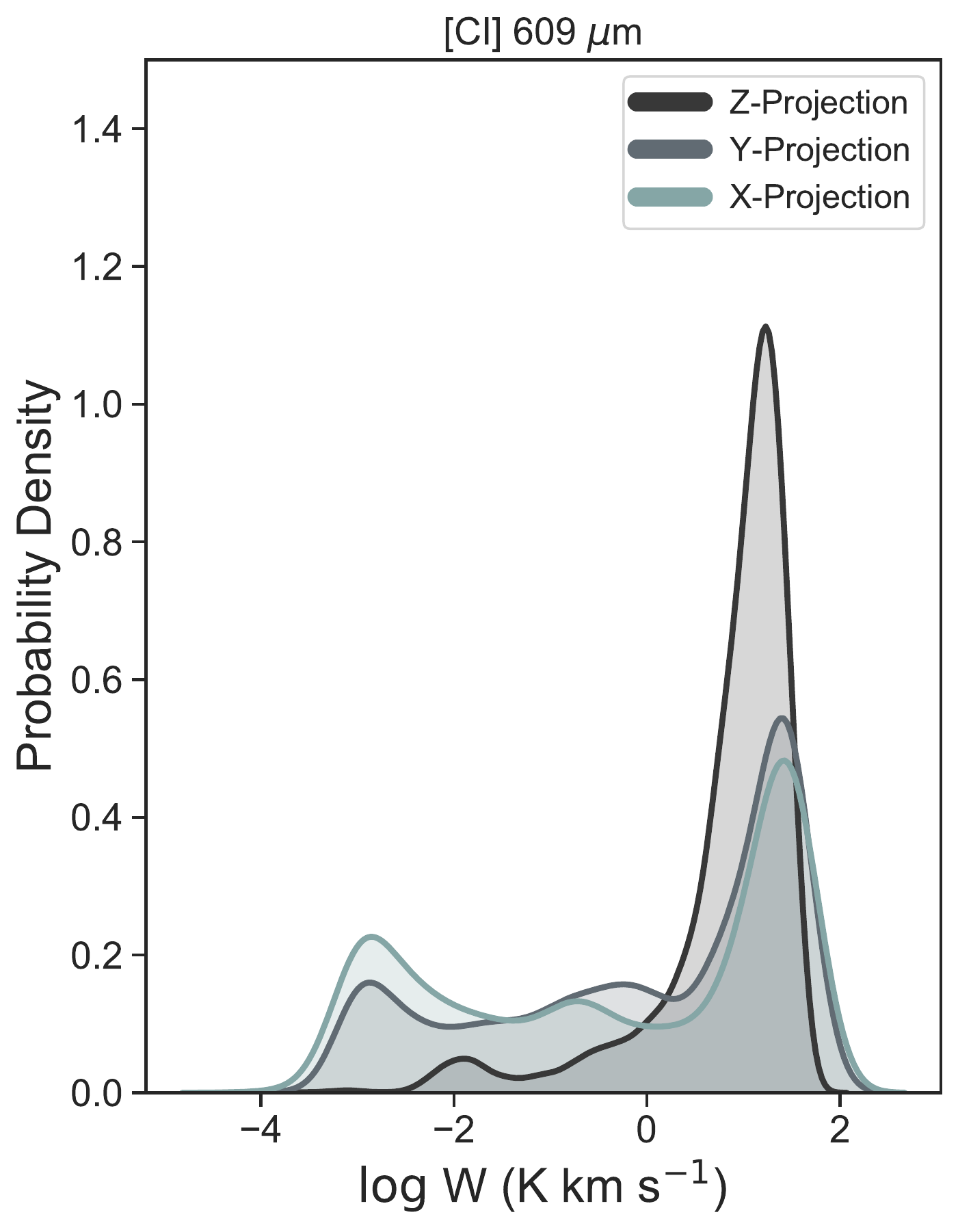}
    \caption{\label{fig:flxDistProjs} PDFs of the velocity integrated emission of [C{\sc i}] 609$\mu$m for three different lines of sight for the $\zeta(N)$ model.}
\end{figure}

\begin{figure*}
    \centering
    \begin{subfigure}[b]{0.3\textwidth}
        \centering
        \includegraphics[width=\textwidth]{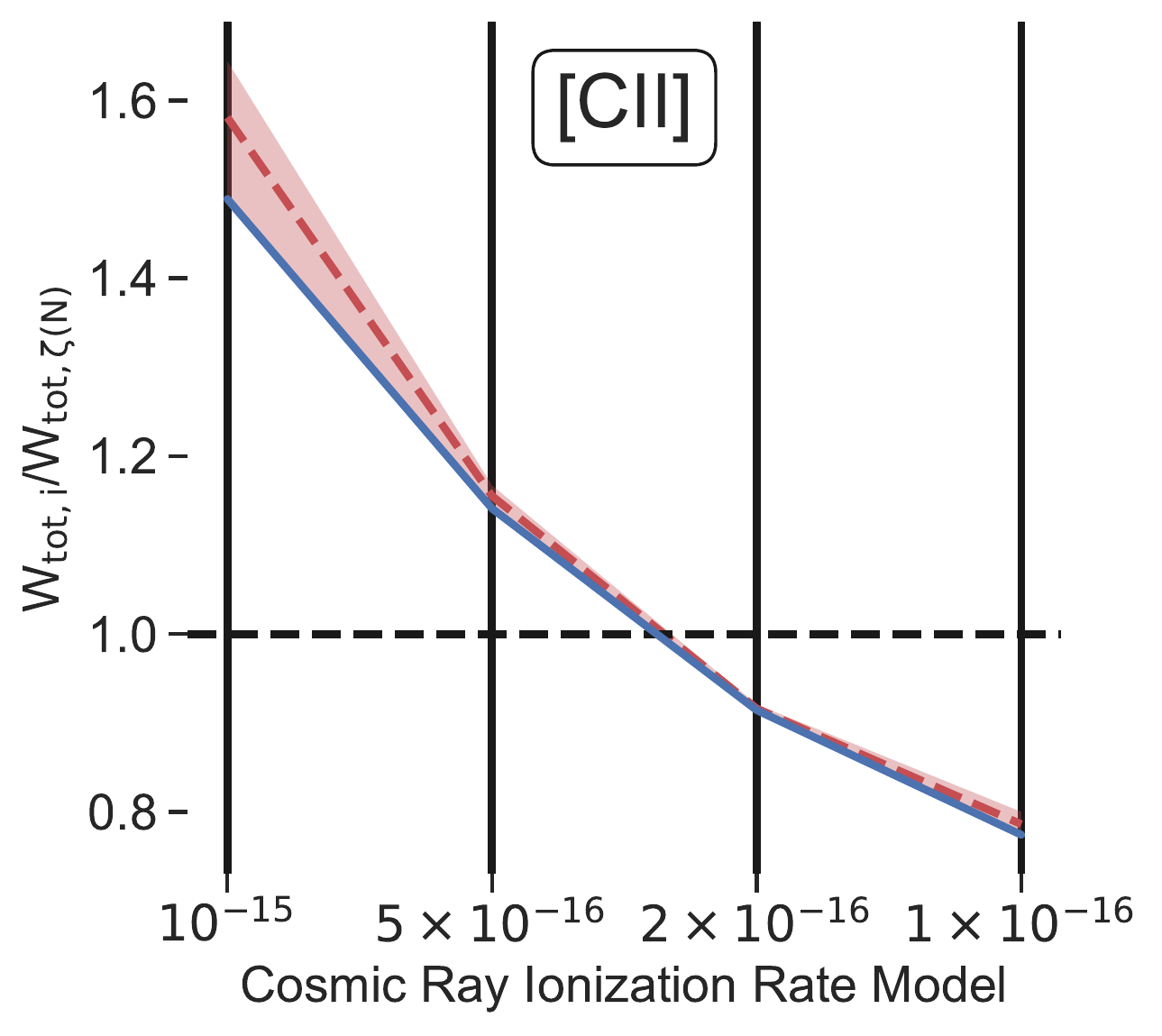}
    \end{subfigure}
    \begin{subfigure}[b]{0.3\textwidth}
        \centering
        \includegraphics[width=\textwidth]{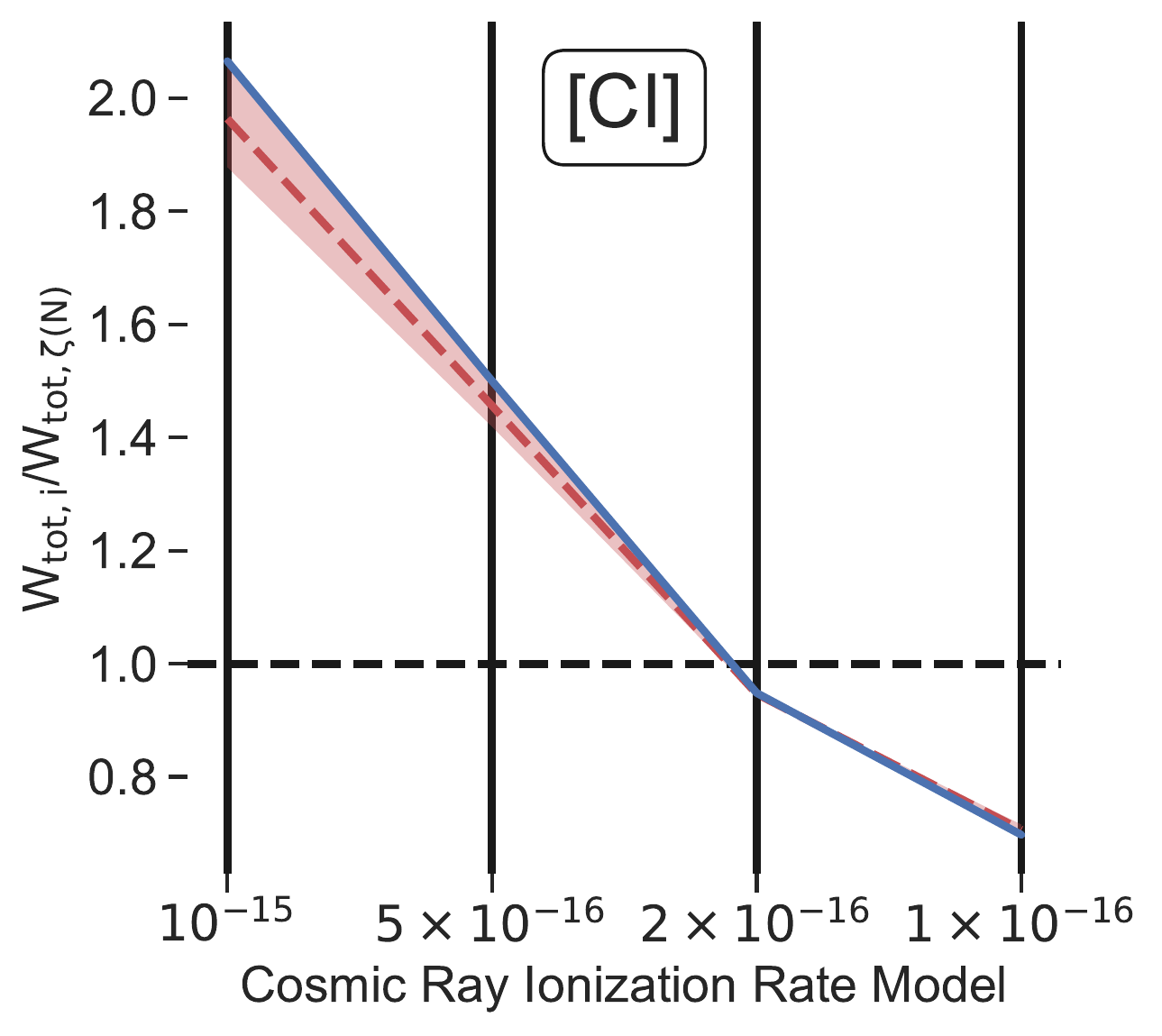}
    \end{subfigure}
    \begin{subfigure}[b]{0.3\textwidth}
        \centering
        \includegraphics[width=\textwidth]{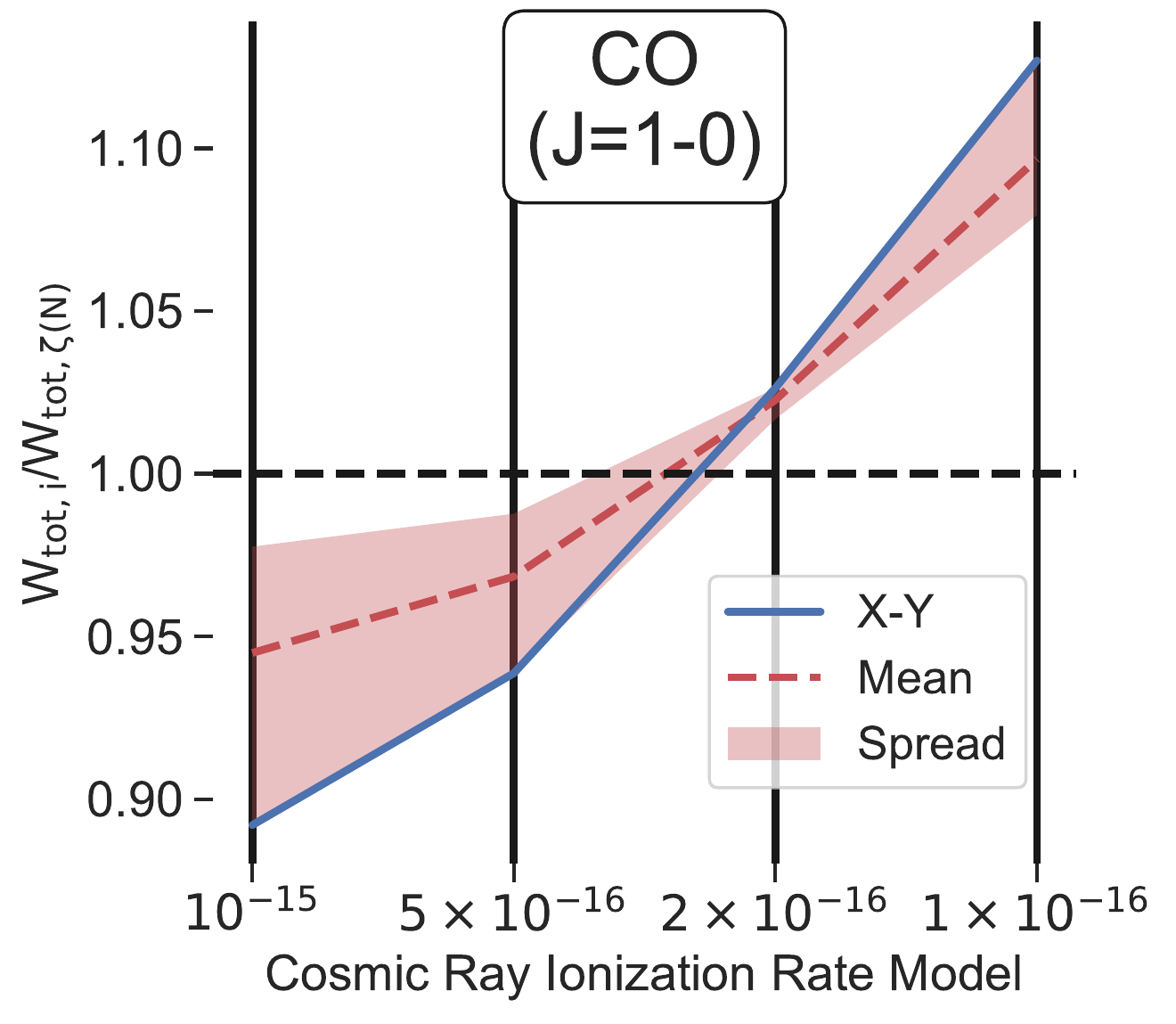}
    \end{subfigure}
    \caption{\label{fig:flxProjErr} Ratios of the total integrated flux from different lines of sight for each of the CRIR models with respect to the $\zeta(N)$ model for [CII] (left), [CI] (middle) and CO (J = 1-0) (right). The blue line denotes the ratio for the Z projection and the red band denotes the overall spread over the three lines of sight.}
\end{figure*}

We now investigate the impact of CR attenuation on the velocity integrated line emissions of [C{\sc ii}] 158$\mu$m, [C{\sc i}] $^3P_1\rightarrow\,^{3}P_0$ at $609\mu$m and the $^{12}$CO $J=1-0$ (hereafter `CO (1-0)') rotational transition at $115.27\,{\rm GHz}$. Figure \ref{fig:intStamp} shows the [C{\sc i}]~$609\mu$m and CO (1-0) integrated emission and the relative errors of these with different constant CRIRs compared to the $\zeta(N)$ model. 

We find that the $5\times10^{-16}$ s$^{-1}$ model is overly bright in [C{\sc ii}]~$158\mu$m and [C{\sc i}]~$609\mu$m while $10^{-16}$ s$^{-1}$ is too dim. For [C{\sc i}]~$609\mu$m there is a low-density pocket in which these trends are flipped.  For CO (1-0), the $\zeta_c = 5\times10^{-16}$ s$^{-1}$ model is too dim in diffuse gas regions, while being over-bright in dense gas (compared to the $\zeta(N)$ model). This trend is flipped in the $\zeta_c = 10^{-16}$ s$^{-1}$ model.

Figure \ref{fig:flxDist} shows probability density functions (PDFs) of the distribution of the emission of [C{\sc ii}]~$158\mu$m, [C{\sc i}]~$609\mu$m and \ce{CO}~(1-0). For [C{\sc i}]~$609\mu$m, the distribution at the low-end is nearly identical, while there are significant deviations for bright regions. The peak brightness changes drastically with different constant ionization rates, highlighting the issues with modeling C chemistry and emission with a constant ionization rate. However, the distributions of the [C{\sc i}]~$609\mu$m emission for the $\zeta(N)$ and $\zeta_c = 2\times10^{-16}$ models are nearly identical. This trend is matched by the \ce{CO} emission, where there is a broad distribution of high emission which is very sensitive to the constant CRIR chosen. For the [C{\sc ii}]~$158\mu$m emission, none of the constant CRIR models well represents the distribution in the $\zeta(N)$ model, although the mass-weighted $\zeta_c = 2\times10^{-16}\,{\rm s}^{-1}$ best matches both the peak and the distribution.

The physical reason behind the changing [C{\sc i}]~$609\mu$m emission peaks lies in the contrast between the gas temperature and the excitation temperature. The decrease in the emission peak is due to the dense-gas ($n_{\rm H}>10^3\,{\rm cm}^{-3}$) temperature dropping below $h\nu/k_{\rm B}\simeq23.1\,{\rm K}$, the temperature associated with the energy of the [C{\sc i}] forbidden-line emission at $609\mu$m (see Appendix \ref{app:one}). Thus, choosing a CRIR to match the diffuse gas will produce higher temperatures in the dense gas. t is worth reminding that the temperatures calculated here are not self-consistent with the \citet{hu2021} results. Their CRIR is scaled linearly with the local star-formation rate, without attenuation included.

Our current analysis has focused on a single projection, along the Z-axis, of the cloud. Here, we evaluate the impact of different lines of sight. In Figure \ref{fig:flxDistProjs}, we show the [CI] 609$\mu$m integrated flux distribution for different projections Figure \ref{fig:flxDistProjs} shows that the distributions are relatively similar, although there is more dim emission in the other projections. The X- and Y-projections are nearly identical. Figure \ref{fig:flxProjErr} shows the ratio of the total integrated flux for each $\zeta_c$ model with respect to the $\zeta(N)$ model for the three emission lines considered here. The blue line shows the Z-projection emission, while the red shade regions shows the overall spread with the line of sight. For the optically-thin emission, the trend of the total flux ratio for different $\zeta_c$ values is the same. For CO, there is greater spread, although the qualitative trends are the same with different lines of sight. The higher spread is expected due to the optical thickness of the line.

\begin{figure}
    \centering
    \includegraphics[width=0.45\textwidth]{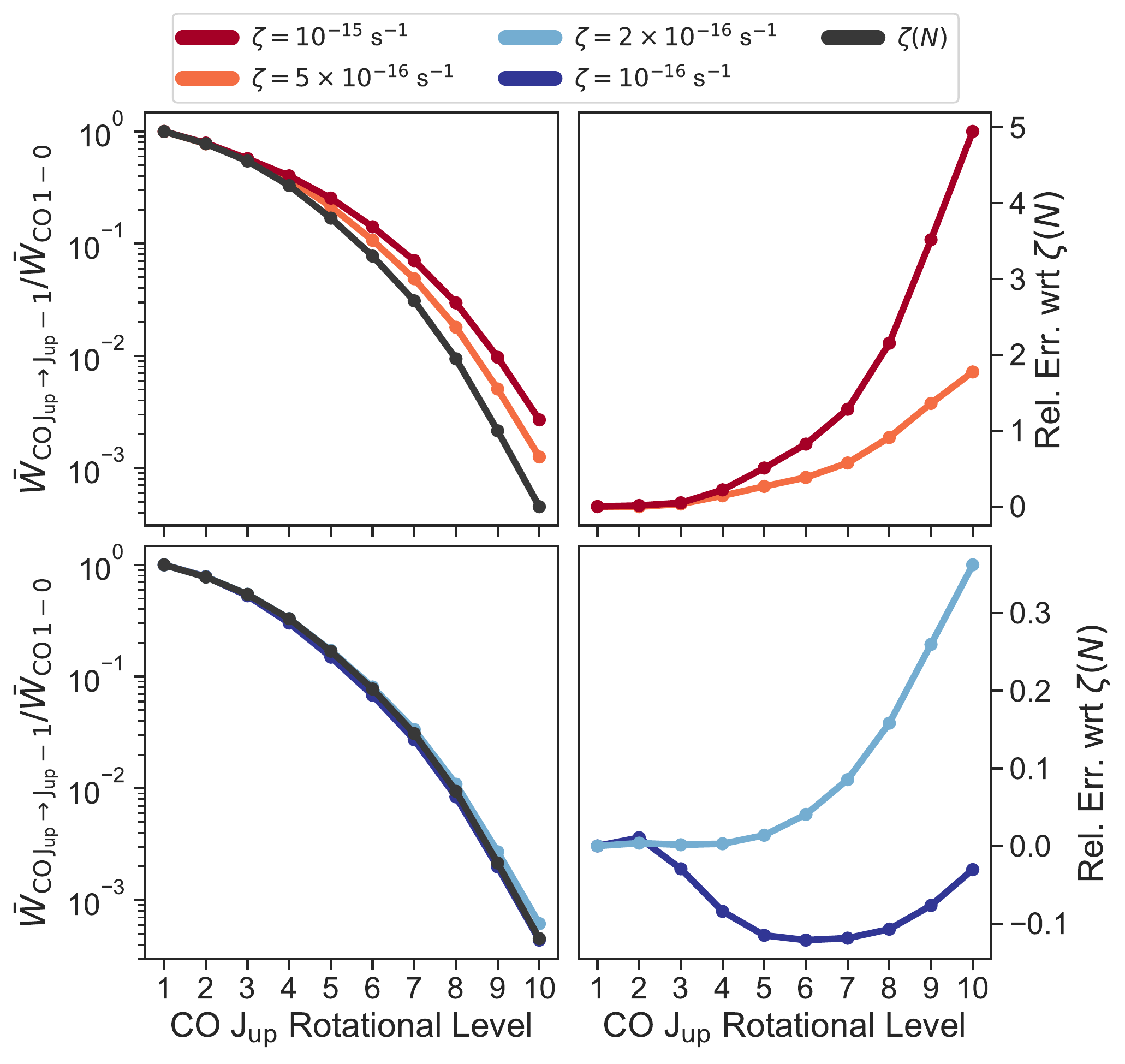}
    \caption{\label{fig:cosled} Left: CO line ratios between higher excitation transitions and the ground transition for the different cosmic-ray ionization rate models. Right: Relative error in the CO line ratios for each $\zeta_c$ with respect to the $\zeta(N)$ model.}
\end{figure}

Finally, we show in Figure \ref{fig:cosled} the \ce{CO} Spectral Line Energy Distribution (SLED) for the different CRIR models. We find that for low-$J$ transitions, all the $\zeta_c$ models well reproduce the line ratios. However, towards higher-$J$ transitions, the differences become more discrepant, due to the attenuation of the CRs resulting in different temperature distributions in dense gas. The $\zeta_c=10^{-16}$ and $2 \times 10^{-16}$ s$^{-1}$ models give the best agreement with the $\zeta(N)$ model, while the higher $\zeta_c$ models produce too much high J-CO  excitation. This is because these lower CRIR models have a mass-weighted mean temperature that is quite similar to the one seen in the $\zeta(N)$ model (see Appendix \ref{app:one}).

\section{Conclusions}

We have presented an analysis of a post-processed photo-dissociation region model of a dense molecular cloud in which we have included a prescription of the CRIR as a function of the local hydrogen column density, $\zeta(N)$. We find the following:

\begin{enumerate}
    \item The column density projection maps of the carbon cycle species, \ce{C+}, \ce{C} and \ce{CO}, are significantly influenced by the inclusion of cosmic-ray attenuation versus utilizing a constant ionization rate. The relative error in the column densities computed using a constant ionization rate versus an attenuated model can be greater than unity (Figure \ref{fig:colStamp}). However, for deeply embedded regions, the impact on the CO column density is marginal.
    \item The choice of CR model has a profound impact on the resulting volumetric abundances of the carbon cycle species, $n_{\rm i}$ (Figs. \ref{fig:ciiabund} -- \ref{fig:coabund}). This effect is particularly strong for \ce{C+} and \ce{C}, with variations greater than an order of magnitude for number densities, $n_{\rm H} > 10^3$ cm$^{-3}$.
    \item We demonstrate that the overall velocity integrated emission of many species is changed by imposing a column density dependent CRIR model, not only due to changes in the abundances, but also in the temperatures of the dense gas ($n_{\rm H}>10^3\,{\rm cm}^{-3}$) because of the volumetric nature of cosmic ray heating. In particular, we find that, irrespective of the chosen line of sight, high constant ionization rates over-produce [CII] and [CI] emission and under-produce CO (J = 1-0) emission, while the inverse is true for low constant ionization rate models. No constant ionization rate model is able to reproduce all lines simultaneously, although the rate, $\zeta_c = 2\times10^{-16}$ s$^{-1}$ has the smallest relative errors for all projections.
    \item For CO, the spectral line energy distribution is also impacted by the chosen CRIR model (see Figure \ref{fig:cosled}). For low-$J$ transitions, $J_{\rm up} < 3$, the total integrated fluxes are similar. However, for high-$J$ transitions, there is a significant variation. Therefore, a CRIR chosen to model diffuse gas may misrepresent the emission in the dense gas.
    \item We showed that the `dense' and `diffuse' clouds have similar distributions of the $\zeta - n_{\rm H}$ phase space. Under the assumption of a general $A_{\rm V, eff} - n_{\rm H}$ relationship, the resulting $\zeta(n_{\rm H})$ function well-reproduces the average trends shown in the $\zeta - n_{\rm H}$ distribution (see Eq. \ref{eq:polyn} and Figure \ref{fig:zeta}).
\end{enumerate}

In summary, we have investigated the effects of CR attenuation on the carbon-cycle chemistry, and compared them to constant CRIR models.
The constant CRIR models cannot fully reproduce the behavior of the attenuated model over all regimes of densities, temperatures and species.
For the $c_k$ coefficients of Table~\ref{tab:crparams} used \citep[][]{padovani2018}, the best match is found for $\zeta_c=2\times10^{-16}$ s$^{-1}$, which reproduces reasonably well the projected emission maps and columns.
However, even this model is not well reproducing the local volume densities in the simulation at high densities.
Therefore, we recommend the following CRIR prescriptions, in descending order of emphasis:
\begin{enumerate}
    \item If local (effective) column densities are available to utilizing a column-dependent prescription for the CRIR (e.g. Eq. \ref{eq:poly}), or,
    \item If local column densities are not available, utilizing the analytic prescription in Eq. \ref{eq:polyn} for the necessary hydrogen-nuclei density for Milky Way-like environments, or
    \item If neither of the above are possible, to utilize the constant CRIR, $\zeta_c = 2\times10^{-16}$ s$^{-1}$, since it resulted in the smallest relative errors in the integrated line emission.
\end{enumerate}

\begin{acknowledgements}
The authors thank the anonymous referee for their comments which improved the clarity of this work. The following {\sc Python} packages were utilized: {\sc NumPy} \citep{numpy}, {\sc SciPy} \citep{scipy}, {\sc Matplotlib} \citep{matplotlib}, {\sc Datashader}, {\sc cmocean}. 
BALG acknowledges support by the ERC starting grant No. 679852 ‘RADFEEDBACK’ and the Deutsche Forschungsgemeinschaft (DFG) via the Collaborative Research Center SFB 956 “Conditions and Impact of Star Formation”. TGB acknowledges support from Deutsche Forschungsgemeinschaft (DFG) grant No. 424563772.
SB acknowledges support from the Center for Theory and Computations (CTC) at the University of Maryland. 
\end{acknowledgements}

\bibliographystyle{aa}
\bibliography{lib} 

\begin{appendix}
\section{Dense-gas temperature distributions}\label{app:one}
\begin{figure}[h!]
    \centering
    \includegraphics[width=0.45\textwidth]{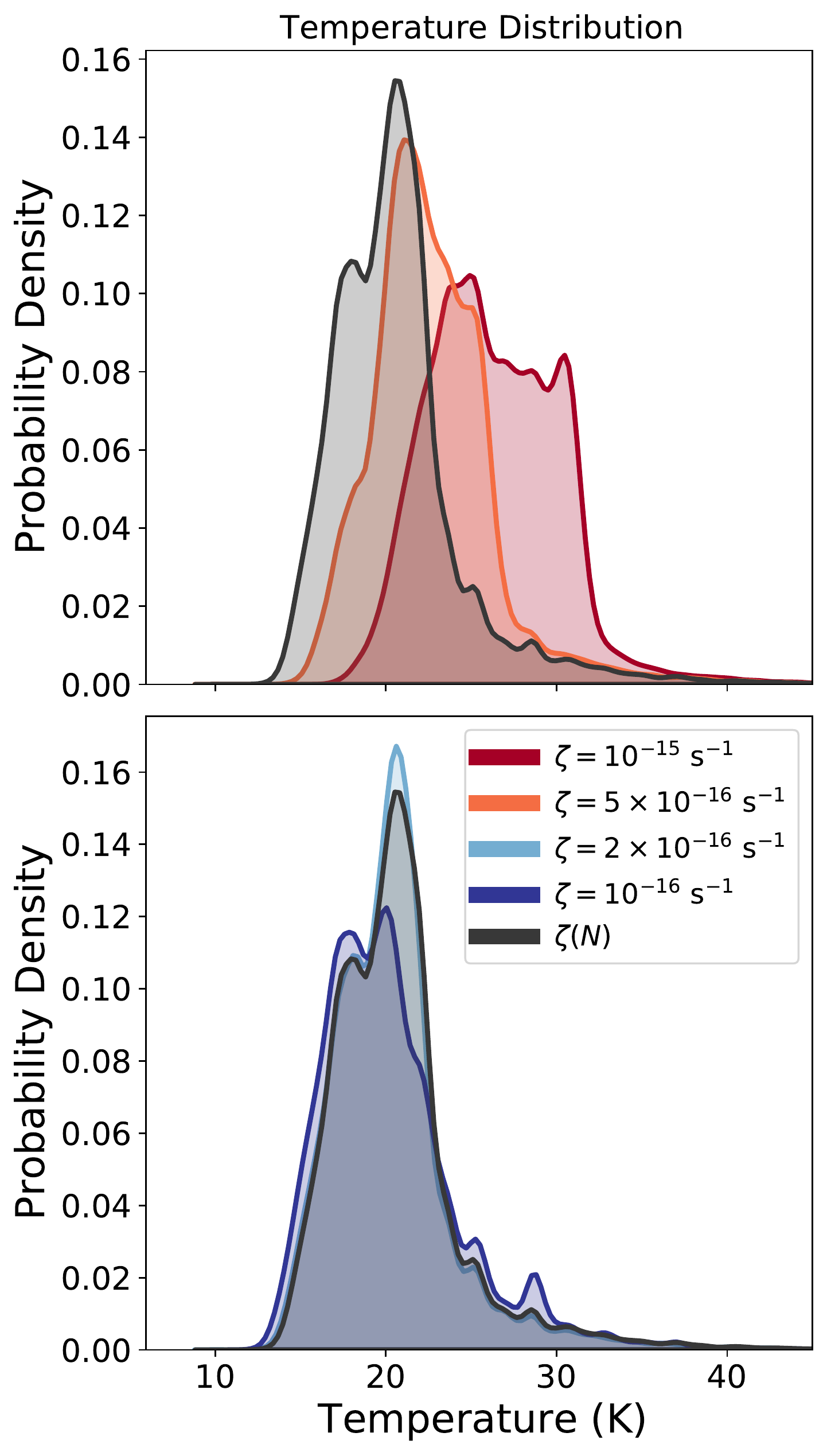}
    \caption{\label{fig:denseT} Temperature PDFs in the dense gas (all cells with $n_{\rm H}{\ge}10^3$ cm$^{-3}$) for the different CRIR models.}
\end{figure}

We investigate the response of the gas temperature in the $n_{\rm H}{\ge}10^3\,{\rm cm}^{-3}$ regime with increasing $\zeta$. Figure~\ref{fig:denseT} shows temperature PDFs for the various simulations including all cells with densities $n_{\rm H}{\ge}10^3\,{\rm cm}^{-3}$. As expected, the distributions follow the trend that higher constant ionization rates produce warmer gas. However, only the $\zeta_c = 2\times10^{-16}$ model reproduces the dense gas temperature distribution of the $\zeta(N)$ model. For observational tracers of the dense gas, the chosen ionization rate may have significant impact on the model emission. Table \ref{tab:tvszeta} shows the mass-weighted gas temperature, $\langle T\rangle_{M}  = \left ( \sum_i T_{\rm gas, i} n_{\rm H,i} \right ) / \sum_i n_{\rm H,i} $ (again, only including cells with $n_{\rm H}{\ge}10^3\,{\rm cm}^{-3}$). We find that the $\zeta_c = 10^{-16}\,{\rm s}^{-1}$ and the $2\times10^{-16}\,{\rm s}^{-1}$ models have $\langle T\rangle_{M}$ closer to the $\zeta(N)$ model. Indeed, the CO SLED for these two models best matches the one of the $\zeta(N)$'s SLED (see \S 3.2).

\begin{table}[h!]
    \caption{Mass-weighted gas temperatures for different cosmic-ray ionization rate models}
    \label{tab:tvszeta}
    \centering
    \begin{tabular}{c|c c}
        CRIR Model & $\langle T \rangle_{M}$ (K)\\
        \hline
                $\zeta(N)$ & 20.38\\
                        $\zeta_c = 10^{-16}$ s$^{-1}$ &  20.24 \\
        $\zeta_c = 2\times10^{-16}$ s$^{-1}$ & 20.79 \\
        $\zeta_c = 5\times10^{-16}$ s$^{-1}$ & 22.64 \\
        $\zeta_c = 10^{-15}$ s$^{-1}$ & 25.40 \\
    \end{tabular}
\end{table}

We highlight the impact of these small changes within the 20-30~K temperature regime that have on higher-$J$ CO transitions for $n_{\rm H}{>}10^3\,{\rm cm}^{-3}$. Table \ref{tab:radex} shows radiation temperature (the equivalent blackbody temperature of the emission in the Rayleigh-Jeans limit\footnote{{\sc Radex} defines this as the main beam antennae temperature with a beam filling factor of unity.}) calculations using {\sc Radex} \citep{radex} of the CO ladder (up to $J=10-9$) for a gas density of $n_{\rm H} = 10^4$ cm$^{-3}$ with different gas temperatures, $T = 20$ and $26$~K, and with $N({\rm CO}) = 10^{18}$ cm$^{-3}$. The aforementioned values are chosen to match with the conditions in the high-density regime as can be seen in Figs.\ref{fig:sim_grid} and \ref{fig:colStamp}. Despite the relatively small difference in the average gas temperature, it has a profound impact on the predictions for the high-$J$ transitions, with deviations up to ${\sim}40$ times for the $J=10-9$ transition. Thus, accurately modeling the temperature of the dense gas is of paramount importance in constraining the physical ISM environment from high-$J$ CO transitions.
\begin{table}[h!]
    \caption{{\sc Radex} model results for the radiation temperature, $T_R$ (K), of the CO J-ladder for $n_{\rm H} = 10^4$ cm$^{-3}$.}
    \label{tab:radex}
    \centering
    \begin{tabular}{c|c c c c c c c c c c}
        Temperature (K) & 1 -- 0 & 2 -- 1 & 3 -- 2 & 4 -- 3 & 5 -- 4 & 6 -- 5 & 7 -- 6 & 8 --
        7 & 9 -- 8 & 10 -- 9\\
        \hline
        20 & $16.44$ & $14.70$ & $12.64$ & $10.56$ & $7.91$ & $3.712$ & $0.77$ & $5.59\times10^{-2}$ & $2.13\times10^{-3}$ & $6.23\times10^{-5}$ \\
        26 & $22.30$ & $20.47$ & $18.24$ & $15.90$ & $13.07$ & $8.41$ & $3.03$ & $5.55\times10^{-1}$ & $4.34\times10^{-2}$ & $2.40\times10^{-3}$ 
    \end{tabular}
\end{table}

\end{appendix}

\end{document}